\documentclass[12pt]{iopart}

\usepackage{amsfonts}
\usepackage{tikz}
\usepackage{graphicx}
\usepackage{url}
\usepackage{fancyvrb}
\usepackage{color}
\usepackage{hyperref}
\usepackage{float}
\usepackage{textcomp}
\usepackage{todonotes}
\usepackage{cite}

\usepackage{graphicx}
\usepackage{color}

\usepackage{xspace}

\newcommand{\degree}{\ensuremath{^\circ}\xspace}

\newcommand{\htp}{\texorpdfstring{\ensuremath{\mathrm{H}_2^+}\xspace}{H2+}}
\newcommand{\hthp}{\texorpdfstring{\ensuremath{\mathrm{H}_3^+}\xspace}{H3+}}

\newcommand{\BE}[0]{\begin{equation}}
\newcommand{\EE}[0]{\end{equation}}
\newcommand{\BEA}[0]{\begin{eqnarray}}
\newcommand{\EEA}[0]{\end{eqnarray}}

\newcommand{\nuebar}{\ensuremath{\bar{\nu}_e}\xspace}

\mathchardef\mhyphen="2D

\newcommand{\figref}[1]{FIG.~\ref{#1}}
\newcommand{\tabref}[1]{TAB.~\ref{#1}}

\newcommand{\opal}{\textsc{OPAL}\xspace}

\newcommand{\opalcycl}{\textsc{OPAL-cycl}\xspace}

\newcommand{\DD}{DAE$\delta$ALUS\xspace}

\newcommand {\mbf}[1]{\mathbf #1}
\newcommand {\RM}[1]{\mathrm{#1}}

\begin{document}

\title{Order-of-Magnitude Beam Current Improvement in Compact Cyclotrons}

\author{Daniel Winklehner, Janet M. Conrad, Devin Schoen, Maria Yampolskaya}
\address{Massachusetts Institute of Technology, 77 Massachusetts Ave, Cambridge, MA, USA}
\ead{winklehn@mit.edu}

\author{Andreas Adelmann, Sonali Mayani, Sriramkrishnan Muralikrishnan}
\address{Paul Scherrer Institut, 5232 Villigen PSI, Switzerland}

\vspace{10pt}
\begin{indented}
    \item[]May 6, 2021
\end{indented}

\begin{abstract}
There is great need for high intensity proton beams from compact particle accelerators in particle physics, medical isotope production, and materials- and energy-research. 
To address this need, we present, for the first time, a design for a compact isochronous cyclotron that will be able to deliver 10~mA of 60~MeV protons - an order of magnitude higher than on-market compact cyclotrons and a factor four higher than research machines.
A key breakthrough is that vortex motion is incorporated in the design of a cyclotron, 
leading to clean extraction.
Beam losses on the septa of the electrostatic extraction channels stay below 50~W (a factor four below the required safety limit), while maintaining good beam quality.
We present a set of highly accurate particle-in-cell simulations, and an uncertainty quantification of select beam input parameters using machine learning, showing the robustness of the design. This design can be utilized for beams for experiments in particle and nuclear physics, materials science and medical physics as well as for industrial applications.
\end{abstract}


\maketitle

\section{Introduction}
\label{sec:intro}
This paper describes the design and simulations of a 10~mA, 60~MeV/amu compact 
cyclotron that can be mass-manufactured. 
Such a machine would have a transformative effect on multiple fields of fundamental and applied science, including neutrino physics, through the IsoDAR project
\cite{bungau:isodar, adelmann:isodar, abs:isodar}; 
isotope production for medicine and other uses
\cite{schmor:isotopes, alonso:isotopes, waites:isotopes};
materials testing for high radiation environments
\cite{jepeal:mad, jepeal:nimb, zinkle:scripta, vincent, moeslang:ifmif}; 
and as a pre-accelerator for a 10~mA, 800~MeV to 1~GeV cyclotron that can 
be used for Accelerator Driven Systems (ADS)
\cite{ishi:adsr,rubbia:adsr, biarrotte:ads,lisowski:ads}
and particle physics (e.g. the \DD experiment 
\cite{alonso:daedalus, aberle:daedalus, abs:daedalus, calabretta:daedalus, conrad:daedalus}).  
We discuss these motivations below and summarize the applications in \tabref{tab:uses}. This beam intensity is an order of magnitude higher than 60 to 100~MeV cyclotrons on the market \cite{IBA, BEST}, and a factor of four higher than the Paul Scherrer Institute - Injector II cyclotron \cite{grillenberger:psi}.
\begin{figure}[!t]
    \centering
    \includegraphics[width=0.8\textwidth]{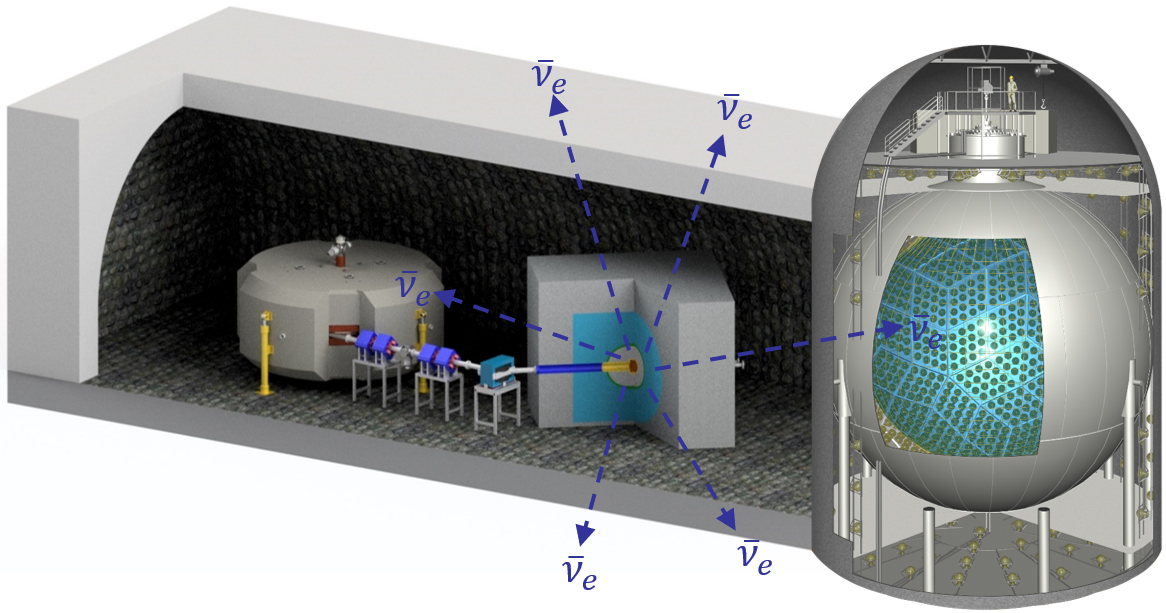}
    \caption{(From \cite{winklehner:mist1}) Schematic of the IsoDAR experiment at Kamioka. 
             From left to right: The cyclotron (ion source on top),
             the medium energy beam transport line,
             the neutrino production target \cite{bungau:target1}, 
             and the KamLAND detector \cite{kunio:kamland}.}
    \label{fig:isodar}
\end{figure}
\begin{table}[!b]
	\centering
    \setlength{\tabcolsep}{4pt}
	\vspace{-10pt}
	\caption{A few potential uses for high current proton beams and how cyclotrons can be 
	         leveraged to reach the goals. 
             ADSR: Accelerator Driven Sub-critical Reactors, 
             ADS: Accelerator Driven Systems for nuclear waste transmutation.
             Cyclotrons can be a cost-effective alternative for tests and demonstrations at the low-power end of the spectrum (tens of mA). \label{tab:uses}}
	
    \vspace{5pt}
    \renewcommand{\arraystretch}{1.25}
    \footnotesize
		\begin{tabular}{llll}
            \hline
            \textbf{Application} & \textbf{Current}  
                                 & \textbf{Energy} & \textbf{Comment}\\
            \hline \hline
            IsoDAR \cite{bungau:isodar, adelmann:isodar, abs:isodar} & 
            10~mA & 
            60~MeV & 
            Use \nuebar from decay-at-rest to search for sterile neutrinos.\\
            
            \DD \cite{alonso:daedalus, aberle:daedalus, abs:daedalus,
                      calabretta:daedalus, conrad:daedalus} & 
            10~mA & 
            800~MeV & 
            A proposed search for leptonic CP violation.\\
            
            ADSR \cite{ishi:adsr,rubbia:adsr} & 
            10-40~mA & 
            $\sim 1$~GeV & 
            Cost-effective alternative for demonstrator experiments.\\
            
            ADS \cite{biarrotte:ads,lisowski:ads} & 
            4-120~mA & 
            $\sim 1$~GeV & 
            Cost-effective alternative for demonstrator experiments.\\
            
            Isotopes \cite{schmor:isotopes, alonso:isotopes, waites:isotopes} & 
            $1-10$~mA & 
            3-70~MeV & 
            Produce more than 250 Ge/Ga generators per week \cite{alonso:isotopes}.\\
            
            Material tests \cite{jepeal:mad, jepeal:nimb, zinkle:scripta, vincent}& 
            10-100~mA & 
            5-40 MeV & 
            Testing of fusion materials similar to IFMIF \cite{moeslang:ifmif}.\\
            \hline
		\end{tabular}
\end{table}

The cyclotron presented here was originally motivated by the need for high-flux sources of neutrinos for the precision study of transformation of neutrino flavor, or oscillations.    This machine was proposed as the first in a two-cyclotron acceleration complex designed for the \DD experiment,  hence it is called the \DD Injector Cyclotron or DIC.  To address the \DD goal of studying $CP$-violation in the neutrino sector \cite{aberle:daedalus}, the complex must produce 10 mA of 800 MeV protons, which, when targeted, results in a well-understood neutrino flux from pion and muon decay-at-rest.   Early on, it was recognized that the DIC also could be used stand-alone, to drive a novel electron anti-neutrino source arising from $^8$Li decay  \cite{bungau:isodar,adelmann:isodar, abs:isodar}.  This concept, proposed as the Isotope Decay-At-Rest experiment (IsoDAR),  targets 10 mA of protons at 60 MeV on beryllium to produce an intense neutron flux that bathes a $^7$Li target producing the required $^8$Li.  The resulting $\beta^+$-decay-produced antineutrino flux allows for tests of $2\sigma$ to $4\sigma$ oscillation anomalies that are attributed to beyond Standard Model particles called ``sterile neutrinos''  \cite{diaz_where_2020}.  IsoDAR can address the sterile neutrino hypothesis at the $>$5$\sigma$ level when the paired with a 1 kton neutrino detector.  The proposed source design, which can be installed in the Kamioka mine in Japan, next to the KamLAND detector \cite{kunio:kamland}, is shown in \figref{fig:isodar}.  The ``IsoDAR cyclotron" and the ``DIC" are identical in design. For consistency, in this paper, we will use ``IsoDAR cyclotron'' 
throughout.


The energy of the IsoDAR cyclotron is similar to cyclotrons proposed for medical isotope 
production \cite{schmor:isotopes}, 
but with an order of magnitude higher beam intensity. 
In particular a modestly-converted IsoDAR cyclotron can produce much-needed 
isotopes ($^{225}$Ac and Ge/Ga generators) for
medical treatment and imaging, as described in two recent publications 
\cite{alonso:isotopes, waites:isotopes}.   
$^{68}$Ge is the parent of the PET imaging isotope
$^{68}$Ga. $^{68}$Ge has a 270 day half-life, making it ideal for
storage and delivery, with the $^{68}$Ga extracted at the hospital.   
One can envision dedicating 10\% of the IsoDAR
running time to production of $^{68}$Ge.  If, instead, a separate version of our cyclotron
is constructed for dedicated isotope production, it can produce more than
250 Ge-Ga generators per week.  
$^{225}$Ac is a valuable alpha-emitter for cancer therapy.  
This new design can impinge 10 mA of protons on a natural thorium target to 
produce, in a dedicated machine, up to 20 doses per hour.    
This would substantially increase the world-wide production rate.

Another example is the use of compact cyclotrons to test materials proposed for 
use inside advanced nuclear reactors and fusion energy devices. Here, 
intense proton beams with energy of 10 to 30~MeV provide a platform to 
achieve relevant materials responses in a fraction of the time compared 
to conventional irradiation methods
inside nuclear reactors \cite{zinkle:scripta, jepeal:mad}. The small footprint and 
relatively moderate costs makes them attractive for university laboratories, 
facilitating student involvement and interdisciplinary research \cite{vincent}.

Among the IsoDAR cyclotron challenges are the strong space charge effects of such a 
high-intensity beam and the small phase acceptance window of an 
isochronous cyclotron, accelerating protons. Space charge,
the mutual Coulomb repulsion of the beam particles inside each bunch, 
matters most in the Low Energy Beam Transport line (LEBT) and during injection 
into the cyclotron. It leads to beam growth and, ultimately, particle loss when
the bunch dimensions exceed the physical constraints of the accelerator.
Phase acceptance poses a similar problem, where particles entering the cyclotron 
at the wrong phase, with respect to the RF cavities' oscillating voltage, will gain
too little or too much energy and consequently go on unfavorable trajectories. 
This leads to energy spread and halo formation. 
Both effects cause overlapping final turns and high particle loss during
extraction, which leads to excess thermal load and activation of the hardware.

To overcome these challenges, the IsoDAR cyclotron concept is based on three innovations:
\vspace{-5pt}
\begin{enumerate}
\setlength\itemsep{0.1em}
    \item Accelerating 5~mA of \htp instead of 10~mA of protons leads to the same 
    number of nucleons on target at half the electrical current, as the remaining 
    electron bound in the \htp molecular ion reduces the electrical current in the beam
    by 50\,\%.
    \item Injecting into the compact cyclotron via a Radio-Frequency Quadrupole (RFQ)
    partially embedded in the yoke aggressively pre-bunches the beam significantly
    increasing the acceptance.
    \item Designing the cyclotron main acceleration to optimally utilize 
    \emph{vortex motion} leads to clean extraction.
    This effect can stabilize beam growth and is explained in Section~\ref{sec:methods}.
\end{enumerate}
\vspace{-5pt}

The focus of this publication lies on the simulation of the main acceleration,
from turn 2 (194 keV/amu) to turn 103 (60 MeV/amu),
and the demonstration of the IsoDAR cyclotron design's capability to 
use vortex motion to keep a stable, round longitudinal-radial bunch shape all the
way to the final turn and thus accelerate 5~mA of \htp while keeping the beam 
losses in the extraction region around 50~W.
An upper limit of 200~W for hands-on maintenance corresponds to 
relative losses on the order of 10$^{-4}$
and is based on practical experience at the Paul-Scherrer-Institute 
Injector II cyclotron \cite{seidel:1.3MW}.
The robustness of the design and simulations is then
shown by means of an uncertainty quantification using machine learning 
techniques.

In the following section, we review the latest IsoDAR cyclotron design.
To give a more complete picture, we also briefly describe the current 
design status of the injection system, including ion source, 
RFQ, and central region, although the particle
distributions resulting from injection system simulations have
not yet been used in the main acceleration simulations.
Methodology and simulation strategies are discussed in Section \ref{sec:methods}.
Our simulations in Section \ref{sec:main} demonstrate that fully 
acceptable beam can be delivered to the extraction system, by careful 
collimator placement, even if the beam is not perfectly matched at injection.
Inclusion of the injection system to close the full start-to-end chain will
likely require retuning of the collimators, but not change our findings.
Finally, Uncertainty Quantification (UQ) using Machine Learning (ML) 
will be shown in Section \ref{sec:uq}.

\section{Hardware Considerations}
\label{sec:design}
\begin{figure}[t!]
    \centering
    \includegraphics[width=0.6\textwidth]{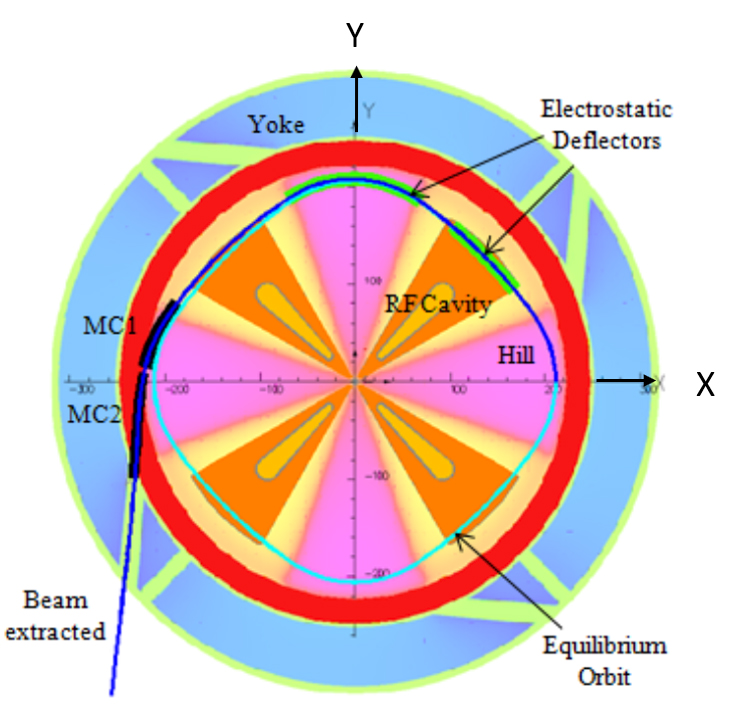}
    \caption{Schematic of the IsoDAR cyclotron.
             Indicated are the hills (magenta) and valleys (yellow) 
             of the isochronous field, the four double-gap RF cavities
             (centered around 45\degree, 135\degree, 225\degree, and 315\degree), 
             the 60 MeV/amu static equilibrium orbit,
             and examples of deflectors and magnetic channels (MC1 and MC2).
             The outer diameter is 6.2 m.
             From \protect\cite{calanna:isodar}.}
    \label{fig:dic_cad}
\end{figure}
This publication focuses on the design, simulation, and uncertainty
quantification of the IsoDAR 60~MeV/amu cyclotron, using, for the first time, 
vortex-motion in the design process.
The IsoDAR cyclotron magnet and RF cavities will be described 
in the following subsection.
The starting point of the simulation study is a particle bunch 
with an average kinetic energy of 193~keV/amu, placed in the first turn.
Because the injection of a high-current beam into a 
compact cyclotron requires care, significant work has also been done on 
the injection system, comprising an \htp ion source, an RFQ 
buncher-accelerator, embedded axially in the cyclotron yoke, 
and the central region of the cyclotron with a spiral inflector.
This RFQ is operated at the cyclotron RF frequency of 32.8~MHz.
A proof-of-concept machine is currently being constructed
to experimentally demonstrate the capability to inject and match the needed 
\htp beam current into the cyclotron: The RFQ-Direct Injection Project (RFQ-DIP)
\cite{winklehner:rfq, winklehner:nima}.
For completeness, we briefly describe RFQ-DIP and 
central region in subsections
\ref{sec:rfq-dip} and \ref{sec:cr}, respectively,
but it should be stressed that
this work is ongoing and a separate publication 
on injection is forthcoming.

\subsection{Cyclotron Magnet and RF Design}
\begin{table}[!b]
    \setlength{\tabcolsep}{4pt}
	\vspace{-10pt}
	\footnotesize
	\caption{Parameters of the IsoDAR cylclotron.
	         Power/cavity assumes a 50\% efficiency at transferring RF power to the beam.}
	\label{tab:dic_params}
	\centering
    \vspace{5pt}
    \renewcommand{\arraystretch}{1.25}
		\begin{tabular}{ll|ll}
            \hline
            \textbf{Parameter} & \textbf{Value} & \textbf{Parameter} & \textbf{Value}\\
            \hline \hline
            $E_{max}$ & 60 MeV/amu	& $E_{inj}$ & 35 keV/amu \\
            $R_{ext}$ &	1.99 m &
            $R_{inj}$ &55 mm  \\
            $\langle B \rangle$ @ $R_{ext}$ &1.16 T	 &	
            $\langle B \rangle$ @ $R_{inj}$ &	0.97 T  \\
            Sectors		& 4		& 	
            Hill width	&	25.5 - 36.5 deg \\
            Valley gap	& 1800 mm	&
            Pole gap	& 80 - 100 mm  \\
            Outer Dia. & 6.2 m	& 
            Full height & 2.7 m  \\
            Cavities	& 4	&
            Cavity type	& $\lambda/2$, 2-gap  \\
            Harmonic & 4$^\mathrm{th}$ &
            rf frequency	& 32.8 MHz  \\
            Acc. Voltage	& 70 - 240 kV	 &
            Power/cavity &	$ 310$ kW  \\
            Coil size & 200x250 mm$^2$ &
            Current dens.	 & 3.167 A/mm$^2$  \\
            Iron weight & 450 tons	&
            Vacuum  & $< 10^{-7}$ mbar \\
            \hline
		\end{tabular}
\end{table}
The IsoDAR cyclotron is a compact isochronous cyclotron operating at 32.8~MHz 
(4$^\mathrm{th}$ harmonic of the particle revolution frequency 8.2~MHz).
The design has been presented in detail elsewhere 
\cite{calanna:isodar, campo:isodar, abs:isodar, calabretta:daedalus}
and will be reviewed here.
A schematic of the machine is shown in \figref{fig:dic_cad} and 
important parameters are listed in \tabref{tab:dic_params}.
\begin{figure}[t!]
    \centering
    \includegraphics[width=1.0\textwidth]{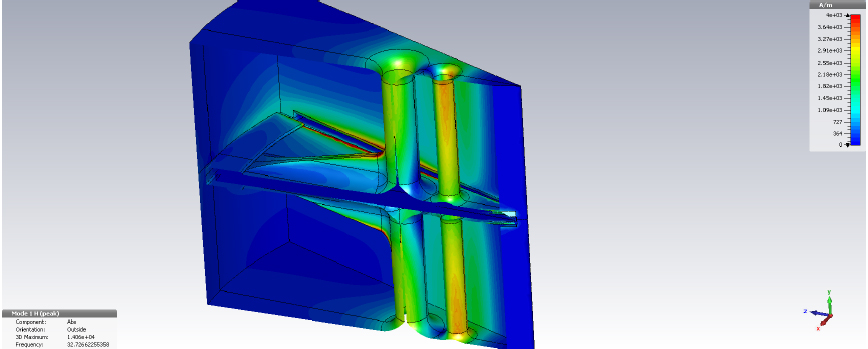}
    \caption{RF cavity modelled in the multiphysics software CST \cite{cst}. The colors correspond to
             surface current density. Central stems are used for support and
             frequency tuning. The cavities are made from 
             oxygen-free high-conductivity (OFHC) copper.
             From \cite{abs:isodar}.}
    \label{fig:rf_cav}
\end{figure} 

Being a compact cyclotron, the main magnetic field is produced by a single pair of
coils encompassing the entire machine and a shared return yoke for the 
four sectors. Notably, the hill gap is large at 100~mm (except for the center, 
where it is reduced to 80~mm to increase vertical focusing) to allow for a 
large vertical beam size. Furthermore, the magnet design includes a $\nu_r = 1$ resonance
crossing close to extraction, which leads to precessional motion and 
improved turn separation. This is achieved
by shaping the magnet poles at larger radii accordingly. 
To increase azimuthal field variation (flutter), 
and thus vertical focusing in the first turn, a vanadium-permendur 
(VP) insert is envisioned on the inner pole tips (see \figref{fig:aima_cad}).
The magnet was designed using the Finite Elements Analysis software OPERA 
\cite{opera} and the generated field was exported for the simulations.

As was shown by Joho \cite{joho:intensity}, a high energy gain
per turn is crucial to acceleration of a high current beam. In the IsoDAR cyclotron,
we place four $\lambda/2$ double-gap RF cavities in the four magnet valleys. 
Their design is based on that of commercial cyclotrons and they are tuned for 
4$^\mathrm{th}$ harmonic operation. 
These RF cavities have a radial voltage distribution going from 70~kV at the 
injection radius to 240~kV at extraction. With a synchronous phase 
$\Phi_{S} = -2.5\degree$ (close to the crest), this amounts to energy gains per 
turn between 500~keV and 2~MeV during the acceleration process. 
The cavity is shown in \figref{fig:rf_cav}. 
The radial voltage distribution calculated in the multiphysics software CST 
\cite{cst} is used in the simulations for each of the eight acceleration gaps.

\subsection{The RFQ-Direct Injection Project}
\label{sec:rfq-dip}
\begin{figure}[t]
    \centering
    \includegraphics[width=1.0\textwidth]{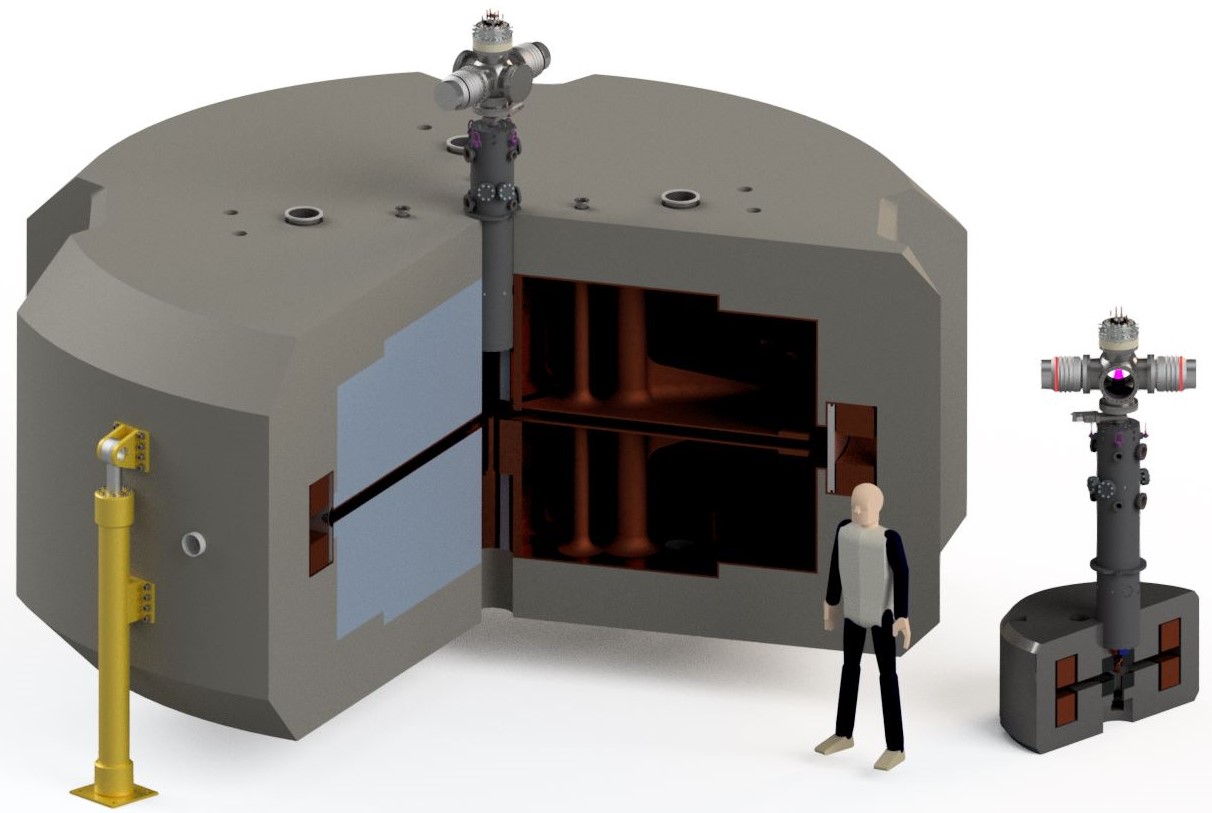}
    \caption{Schematic of the IsoDAR cyclotron (left) next to the 
             RFQ Direct Injection Prototype (right). 
             Ions are produced in the ion source (top), are accelerated and 
             bunched in the RFQ (middle) and injected into the cyclotron 
             central region to be accelerated.}
    \label{fig:rfq-dip}
\end{figure} 
RFQ-DIP \cite{winklehner:rfq, winklehner:nima} is the prototype of a 
novel injection system for compact cyclotrons. 
A cartoon of the device is shown in \figref{fig:rfq-dip}.
RFQ-DIP comprises a multicusp ion source (MIST-1) \cite{axani:mist1,
winklehner:mist1}, a short matching LEBT with chopping and steering
capabilities, and an RFQ that is embedded in the cyclotron yoke, 
to axially inject a highly bunched beam into the central region
through a spiral inflector. 
By aggressively pre-bunching the beam, we fit more particles into the RF phase acceptance window of $\approx20\degree$.
The system is designed to produce and inject up to 15~mA of \htp.
Early commissioning runs with MIST-1 have shown a 76\,\% \htp fraction
at 11~mA/cm$^2$ and a maximum current density of 40 mA/cm$^2$ when the
source is tuned for \hthp \cite{winklehner:mist1}.
This is currently a factor 4 short of the design goal. However,
further upgrades to cooling and extraction system are ongoing that
we anticipate will yield the necessary beam currents.
In Ref. \cite{winklehner:nima}, we also describe the physics design of 
the RFQ linear accelerator-buncher that will be embedded in the cyclotron yoke. It will deliver a highly bunched beam to the spiral inflector -- an electrostatic device that bends the beam from the axial direction
into the acceleration plane of the cyclotron (median plane, or mid-plane), where the beam is accelerated and matched to the cyclotron main acceleration (described in Section \ref{sec:main}).
Due to the high bunching factor and strong space charge, the beam starts 
diverging in transverse direction and de-bunching in a longitudinal direction 
soon after the RFQ exit. To mitigate this, we included a re-bunching 
cell in the RFQ design and place an electrostatic quadrupole focusing 
element before the spiral inflector.
Furthermore, the spiral inflector electrodes can be carefully shaped to 
add vertical focusing.
First simulations of the full injector 
(up to the exit of the spiral inflector) showed transmission of
$\approx78\%$ for two test cases (10 mA and 20 mA of total beam current, 
80\% \htp, 20\% protons) with transverse emittances of $0.3-0.4$ mm-mrad 
(RMS, normalized) and longitudinal emittances of $7-8$ keV/amu-ns
(RMS)~\cite{winklehner:nima}.

\subsection{Central region}
\label{sec:cr}
\begin{figure}[b]
    \centering
    \includegraphics[height=0.35\textwidth]{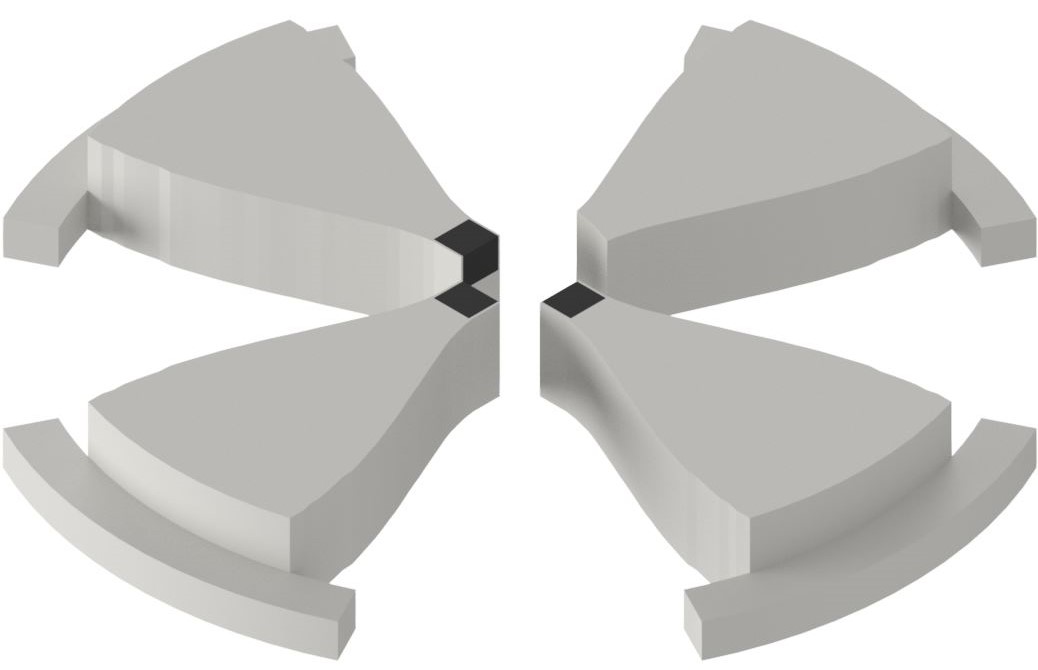}
    \includegraphics[height=0.35\textwidth]{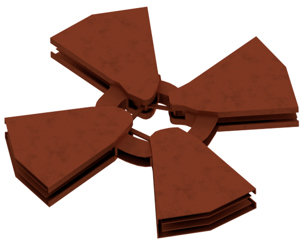}
    \caption{CAD renderings of the central region. Left: The iron poles for 
             magnetic field calculations (lower half only). 
             The VP inserts can be seen in black at the pole tips. 
             One of the pole tips is truncated, yielding space for the spiral
             inflector. Right: The RF electrodes with tips angled to adjust 
             the phase during the first two turns.}
    \label{fig:aima_cad}
\end{figure} 

In parallel with the simulation study presented in this manuscript, and
complementary to it (overlapping in the first four turns of the cyclotron), 
a detailed central region study, subcontracted to the company 
AIMA Developpement in France,
was performed, and summarized in a technical report \cite{aima:central}. 
In this study, a 3D magnetic field was generated that includes the effects 
of VP inserts in the pole tips, and mimics the center field of the 
IsoDAR cyclotron. One pole tip was cut short to make room for 
the spiral inflector (see \figref{fig:aima_cad} (left)).
The VP has a sharper turn in the B-H curve, slightly improving the 
flutter in the central region. An optimized dee electrode system was 
generated, which can be seen in \figref{fig:aima_cad} (right).
This system exhibits good vertical focusing and small orbit center 
precession. The dee peak voltage was increased to 80 kV from the 
nominal 70 kV in the IsoDAR baseline, which is high, but achievable.
Particle distributions from a simulation of the ion source extraction
and RFQ injector (described in the previous section) were used as 
initial conditions in the AIMA study.

\begin{figure}[!t]
    \centering
    \includegraphics[width=0.8\textwidth]{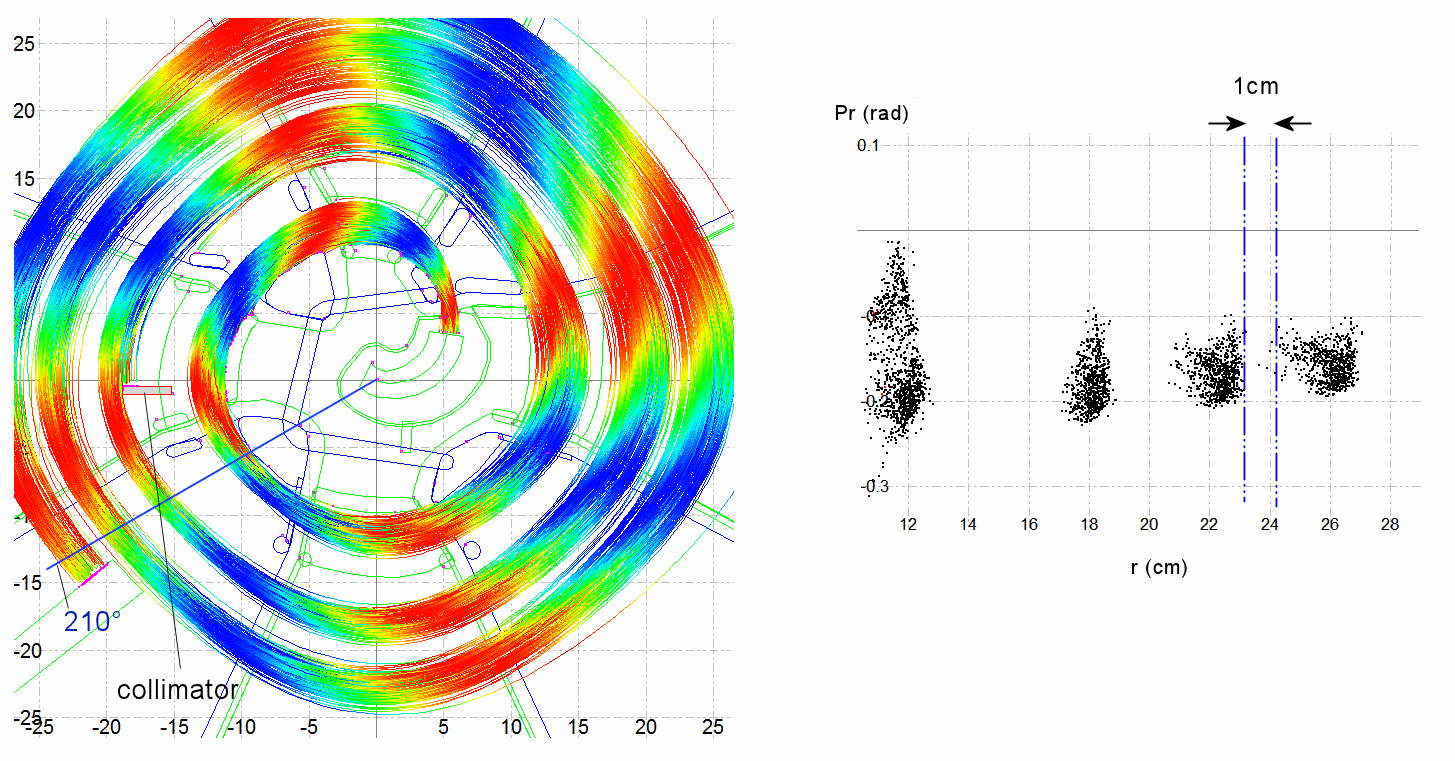}
    \caption{Left: Trajectories of the first 3.5 turns (2 MeV) in the simulated central region.
             Right: Demonstrated turn separation of 1 cm (edge-to-edge) after placing a single collimator
             in the first turn. Beam transmission from the entrance of the spiral inflector
             to the probe was 42\%. From \cite{aima:central}.}
    \label{fig:aima_result}
\end{figure} 

The desired edge-to-edge turn separation of 10~mm in the fourth turn (at 
1~MeV/amu beam energy) was achieved by placing a single collimator in the first turn
(see \figref{fig:aima_result}). 
The combined beam loss in the spiral inflector and in the 
central region (mostly on the single collimator)
was 58\%  and thus the cumulative transmission efficiency was 42\%
from ion source to turn 4 of the cyclotron.
Should we take these results at face value, a total current 
of 20~mA (comprised of 80\,\% \htp and 20\,\% protons) would be 
needed from the ion source and injected into the RFQ. 
In Ref. \cite{winklehner:nima} we showed that the RFQ-DIP system 
can handle such a current. 
However, this central region study, as of yet, does not include space charge effects (space charge was included only up to the entrance of the spiral inflector, cf. previous section).
As we will show in Section \ref{sec:main}, including space-charge
will, somewhat counter-intuitively, improve the situation, as 
the vortex-effect will help maintain a stable distribution and
only halo particles will have to be removed with collimators.
Furthermore, placement of several collimators instead of one allows
more control over which particles are removed, yielding lower losses.
This does not hold for the spiral inflector itself, where including
space charge will lead to slightly lower transmission.
Future work will combine the results presented in Section \ref{sec:main}
with the design work performed in the AIMA study, by importing the 
3D magnetic and electric fields of the CAD model into OPAL
and tracking with space charge,
using the cyclotron injection mode described in Ref \cite{winklehner:spiral}.

\section{Methodology}
\label{sec:methods}

\subsection{OPAL simulation code}
\opal~\cite{adelmann:opal} is a suite of software for the simulation of particle accelerators, which originates at the Paul Scherrer Institute,
and which is programmed in C++.
One of the available flavors is
\opalcycl, which is specifically created to simulate cyclotrons, 
and which we used for this study.
The following is a brief summary of the description in \cite{winklehner:spiral}.
\opal uses the Particle-In-Cell (PIC) method to solve the 
collisionless Vlasov equation 
\begin{equation*}\label{eq:Vlasov}
 \frac{df}{dt}=\partial_t f + \sum_{j=1}^{M} \left [ \frac{\partial f}{\partial \mbf{x}_j} \dot{\mbf{x}}_j + q (\mbf{E}+ c\mbox{\boldmath$\beta$}\times\mbf{B})_j \frac{\partial f}{\partial \mathbf{P}_j} \right ],
\end{equation*}
in the presence of external 
electromagnetic fields and self-fields,
  \BEA
    \label{eq:Allfield}
    \mbf{E} & = & \mbf{E_{\RM{ext}}}+\mbf{E_{\RM{self}}},\\    
    \mbf{B} & = & \mbf{B_{\RM{ext}}}+\mbf{B_{\RM{self}}}.
  \EEA
Here, $\mbf{x}$ and $\mbf{P}$ are the canonical position and momentum of the particles
in the distribution function
\[
f(\mbf{x},\mathbf{P},t): (\Re^{3M} \times \Re^{3M} \times \Re) \rightarrow \Re,
\]
and $M$, $c$, $t$, $q$, and $\mbox{\boldmath$\beta$} = \mbf{v}/c$, the number of simulation particles, 
vacuum speed of light, time, charge of a particle, and velocity scaled by c, respectively.
A 4\textsuperscript{th} order Runge-Kutta (RK) integrator
is used for time integration. External fields are evaluated four times 
per time step. Self-fields are assumed to be constant during one time step, 
because they typically vary much slower than the external fields.

The self fields $\mbf{E_{\RM{self}}}$ and $\mbf{B_{\RM{self}}}$ are calculated on a grid using a Fast Fourier Transform (FFT) method.
The external fields $\mbf{E_{\RM{ext}}}$ and $\mbf{B_{\RM{ext}}}$
can be calculated with any method of the users choosing
and then loaded into \opal either as a 2D median plane field (magnetic field only) 
or a full 3D electromagnetic field map.
OPAL uses a series expansion to calculate off-plane elements from the 
2D median plane fields.
Furthermore, the 3D maps are time-varied according to
\begin{equation*}
\mathbf{E}_\RM{ext, 3D}(t) = \mathbf{E}_\RM{ext, 3D, 0} \cdot \cos(\omega_{\RM{RF}} t - \phi_S)
\end{equation*}
with $\omega_{\RM{RF}}$ the cyclotron RF frequency and $\phi_S$ the phase.
If a static 3D field is desired, the frequency and phase can be set to zero.
Here, we used OPERA to calculate the median plane field and COMSOL \cite{comsol} 
for the 3D electrostatic fields of the extraction system.

\opalcycl comes with a number of built-in diagnostic devices. 
One such diagnostic is the \opal \verb|PROBE|. 
It is a 2D rectangle placed in the 3D simulation space. 
Whenever a particle crosses the probe plane, it is registered and the 
particle data is added to the probe data storage. In Section \ref{sec:main}
we denote probes with a single line in a top-down view of the cyclotron.
Trajectory data, probe data, and data of particles lost on collimators 
are stored by \opal in separate files in HDF5 \cite{hdf5} data format. 
All post-processing is done in Python 3.7 \cite{python}.

\opal has been extensively tested and benchmarked. 
Pertaining to the cyclotron studies presented here, we cite three examples: 

\opal was used to study beam dynamics in PSI Injector II, where high-fidelity
simulations of the full cyclotron using $10^6$ particles
were performed that showed the formation of a stable vortex \cite{yang:vortex}
(cf. Subsection \ref{sec:vortex}).
In \cite{yang:vortex}, the effects of radially neighboring bunches in the PSI
ring cyclotron were also investigated.
A comparison of \opal simulations with radial probe measurements in Injector II, 
yielding good agreement, was shown in \cite[Fig.~3]{seidel:1.3MW}.
More recently, a detailed study was performed for the planned 3~mA upgrade of 
PSI Injector II that further corroborates the fidelity of the code and 
the applicability of the vortex motion design concept for high
intensity cyclotrons \cite{kolano:psi1}.

\subsection{Coordinate systems}
Trajectory data and general layout images are
shown in the laboratory frame (global coordinates). However, as vortex motion
happens through coupling in the longitudinal-radial plane and collimators
can only scrape particles that extend away from the bunch in the direction 
perpendicular to the direction of bunch movement (mean momentum), it is convenient
to look at the bunch in a local frame (local coordinates) that are defined as
follows: 
vertical: $\tilde{\mathrm{z}}$ = z, 
longitudinal: $\tilde{\mathrm{y}}$ = direction of mean bunch momentum, 
transversal (also called ``radial'' here): $\tilde{\mathrm{x}}$ = orthogonal 
to $\tilde{\mathrm{z}}$ and $\tilde{\mathrm{y}}$. 
N.B.: The radial direction does not necessarily coincide with the ray 
originating in the origin and passing through the bunch center, as the 
magnetic field is not uniform, but has hills and valleys.

\subsection{Vortex-motion}
\label{sec:vortex}
\begin{figure}[t!]
    \centering
    \includegraphics[width=0.8\textwidth]{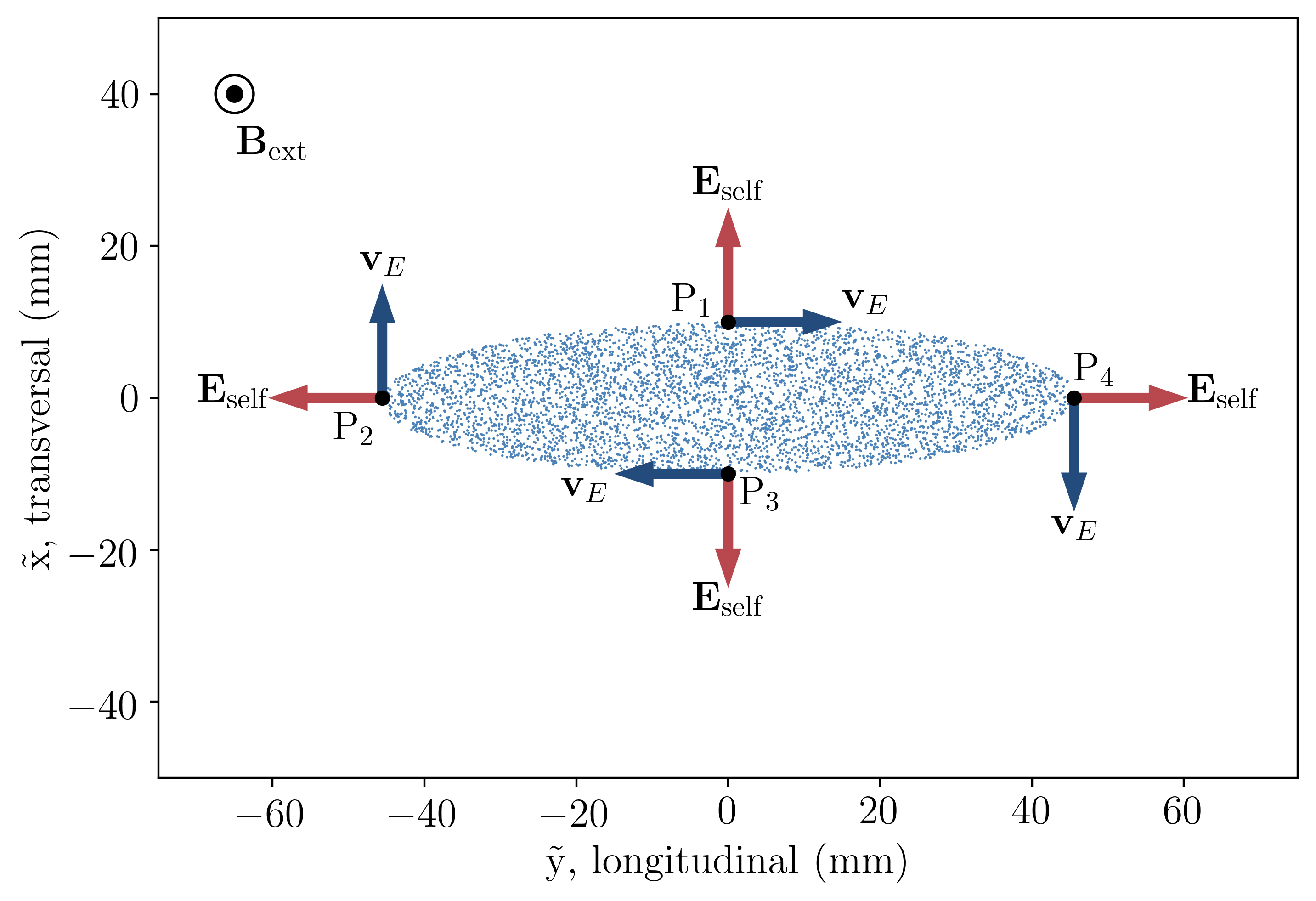}
    \caption{An intuitive picture of vortex motion. 
             The beam is presented in its local frame (cf. text) 
             and the direction of the additional velocity component $\mbf{v}_E$
             due to an $\mbf{E}\times\mbf{B}$ drift is indicated for the four extrema 
             of the bunch. Inspired by \cite{kleeven:vortex}.}
    \label{fig:vortex_intuitive}
\end{figure}
\begin{figure}[t!]
    \centering
    \includegraphics[width=1.0\textwidth]{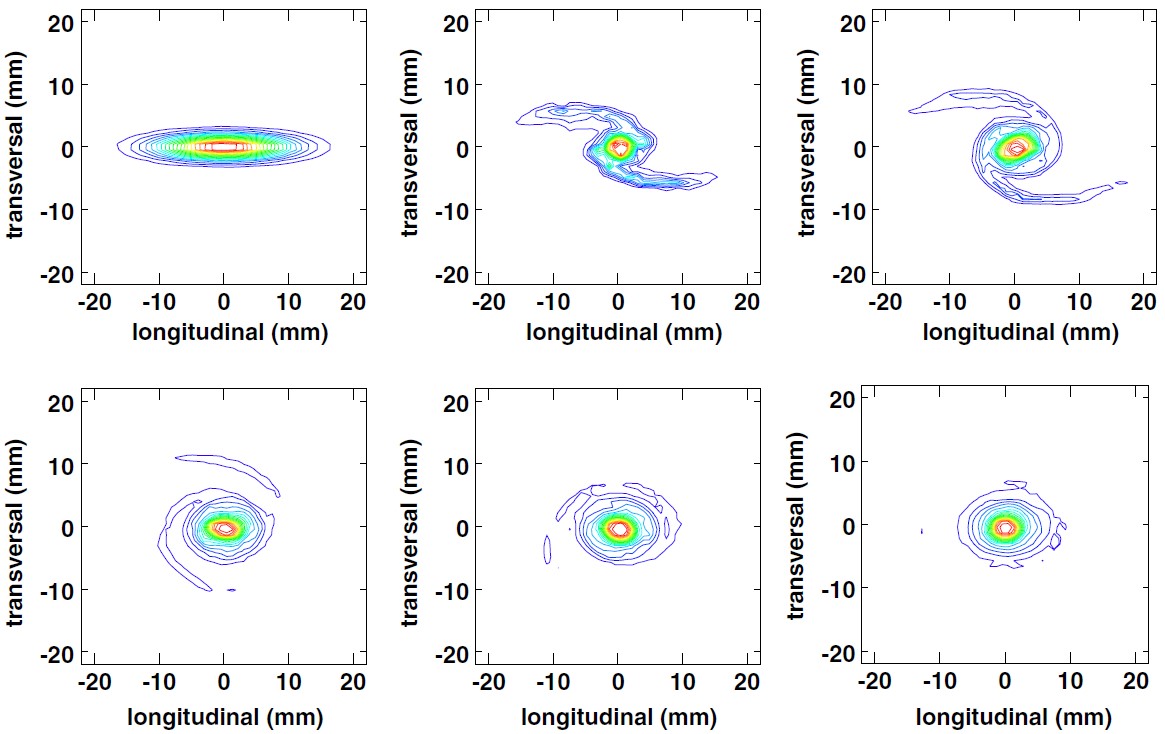}
    \caption{\opalcycl simulation of a single, coasting bunch in PSI injector II,
             shown in its local frame, moving to the left. 
             Beam current: 1~mA, beam energy: 60~MeV.
             Left to right: Upper row: turn 0, 5, 10, 
             lower row: turn 20, 30, 40. From \cite{yang:vortex}.}
    \label{fig:yang_vortex}
\end{figure}
In isochronous cyclotrons, the interaction between the self fields 
of the beam, arising from space charge, and the external magnetic forces,
from the cyclotron main magnet, can lead to the formation of a stable, 
almost round, spatial distribution in the horizontal plane. 
In this subsection, we give a brief overview on the current understanding 
of this effect, dubbed \emph{vortex motion}.

Vortex motion was first seen in PSI Injector II, and subsequently 
investigated and confirmed both experimentally and through computer 
simulations \cite{stetson:vortex, koscielniak:vortex, adam:vortex}.
A simplified, but intuitive picture inspired by Ref 
\cite{kleeven:vortex} is shown schematically in \figref{fig:vortex_intuitive}.
Here, only the force at the four extrema (longitudinal and radial minima and 
maxima) of the bunch in the local frame 
($\tilde{\mathrm{x}}$, $\tilde{\mathrm{y}}$, $\tilde{\mathrm{z}}$) are considered.
This is simply the Lorentz force due to self fields and external fields:
\begin{equation*}
    \mbf{F} = q \cdot (\mbf{v} \times \mbf{B_{\RM{ext}}}) + q \cdot \mbf{E_{\RM{self}}} 
\end{equation*}
with $\mbf{B_{\RM{ext}}} = \mbf{e}_{\tilde{\mathrm{z}}} B_0$. Here $\mbf{e}_{\tilde{\mathrm{x}}}$, $\mbf{e}_{\tilde{\mathrm{y}}}$, and 
$\mbf{e}_{\tilde{\mathrm{z}}}$ are the coordinate vectors of the local frame.
Neglecting for a moment
the self term, the solution to the equation of motion would be 
the usual circular motion of the particles in the magnetic dipole
field. Assuming mid-plane symmetry, $E_{\tilde{\mathrm{z}}}$ must
be zero, and the addition of the self term leads to an 
$\mbf{E}\times\mbf{B}$ drift in the $\tilde{\mathrm{x}}$-$\tilde{\mathrm{y}}$ plane that adds an additional 
velocity term to each particle:
\begin{equation*}
    \mbf{v}_E = \frac{\mbf{E_{\RM{self}}}\times\mbf{B_{\RM{ext}}}}{\mathrm{B_{\RM{ext}}}^2}.
\end{equation*}

For example, at the head of the bunch (P$_1$ in \figref{fig:vortex_intuitive}), 
$\mbf{E_{\RM{self}}} = \mbf{e}_{\tilde{\mathrm{x}}} E_{\tilde{\mathrm{x}}}$ and consequently, 
$\mbf{v}_E = \mbf{e}_{\tilde{\mathrm{y}}}\cdot(E_{\tilde{\mathrm{x}}} B_0)/\mathrm{B}^2$. 
Similar relations hold for P$_2$, P$_3$, and P$_4$ and lead to the velocity 
vectors indicated in \figref{fig:vortex_intuitive}, which in turn lead to the spiraling motion 
in the local frame. A simulation of this effect in PSI Injector II is
shown in \figref{fig:yang_vortex}.
In reality, the situation is, of course, more complex, as the magnetic field
is not uniform but an Azimuthally Varying Field (AVF) and the space charge
force is not linear, as was assumed in the intuitive picture.


An early theoretical approach to vortex motion was presented in 
\cite{bertrand:vortex}. A more rigorous treatment for AVF cyclotrons was presented
in \cite{baumgarten:vortex1} and later extended to the central region
and injection in \cite{baumgarten:vortex2}. These methods allow 
finding matched distributions under the assumption that the energy gain per 
turn is small compared to the beam energy (adiabatic energy gain).
These theories together with simulations and experimental studies suggest that, 
in order for the beam to be well matched, a very short bunch with minimal energy
spread should be injected into the cyclotron at high energy.
In practice, this is not possible in a compact cyclotron with axial injection,
as the spiral inflector typically can only hold voltages $<20$~kV without sparking,
due to space restrictions.
As described at the end of \cite{baumgarten:vortex2}, in the center of the
cyclotron, where energy gain cannot be adiabatic by the nature of the machine, 
a careful collimation process must then be employed to shape the bunch in 
longitudinal-radial phase space. This is what we have done here
and what is described in Subsection \ref{sec:colls}.
At higher energies (main acceleration), the stable distribution has then
formed and all the points of \cite{baumgarten:vortex1} hold, which means that
very high current beams can be accelerated, albeit with significant
losses on the central region collimators to cut away halo until the 
stable round distribution has formed.

\subsection{Collimator modelling}
\begin{figure*}[t!]
    \centering
    \includegraphics[width=1.0\textwidth]{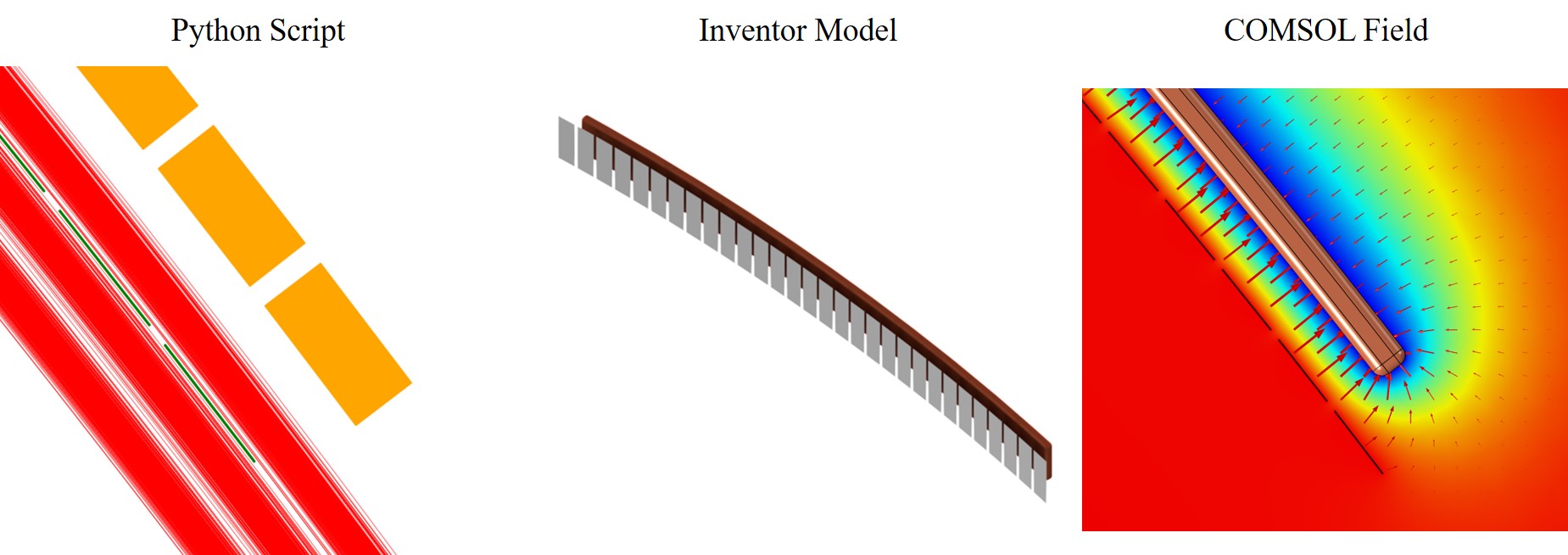}
    \caption{Calculation of the 3D electrostatic septum field in three steps:
             Left: Calculation of the coordinates in Python with visual feedback.
             Center: Export into a Autodesk Inventor macro to generate the 3D CAD model.
             Right: Import into COMSOL and calculation of the field.}
    \label{fig:sep_gen}
\end{figure*}
\label{sec:colls}
Collimators are placed in the \opalcycl code as \verb|CCOLLIMATOR| objects in the
input scripts using the following syntax \cite{OPAL-Manual}:
\begin{verbatim}
    "Name": CCOLLIMATOR, WIDTH=w
            XSTART=x1, XEND=x2, 
            YSTART=y1, YEND=y2,
            ZSTART=z1, ZEND=z2;
\end{verbatim}
where ``Name'' is a unique label for the collimator, x1, x2, y1, y2
are the start and end coordinates along the direction of movement,
w is the width of a single collimator block perpendicular to the direction of
movement (both in the cyclotron median plane), and z1, z2 mark the vertical
extent. \opal terminates particles intersecting with collimators and
saves the particle data of lost particles in an HDF5 file.

The manual optimization of collimator placement is an iterative process
consisting of the following steps:
\begin{enumerate}
    \item The beam is tracked for ten turns while saving the full 
          6D particle distributions 250 times per turn (2500 data-sets).
    \item Particle distributions at each step are projected onto the median
          plane and transformed into their local frame.
    \item Good positions (time steps) to scrape halo particles are manually selected,
          length, width and height are specified by the user,
          and a Python script is used to generate the text to add the new
          collimators to the \opal input file.
    \item Return to Step 1: The simulation is run again with the new collimator(s). 
    \item Occasionally, all 103 turns are simulated while saving particles 
          only 4 times per turn and the data on the probes is analyzed to see 
          what the anticipated beam loss on the septum will be.
\end{enumerate}

The placement of collimators used in Section \ref{sec:main} (cf. \figref{fig:colls})
is optimized by hand, following the process described above. 
Using the surrogate modelling described in Section \ref{sec:uq}, 
an optimization of the radial collimator positions was 
performed that yielded no significant improvement, showing that the solution
is robust.

\subsection{Extraction Channel Modelling}
The electrostatic extraction septa (grounded) and corresponding puller 
electrodes (at negative high voltage potential) are generated in an iterative 
process.
In each iteration, the same workflow is used to obtain a 3D electrostatic
field map for OPAL.
We limit our description to the process for Septum 1 with the understanding that,
with the exception of the azimuthal position of the septum, it is identical to
that for Septum 2.
The workflow in each iteration comprises the following steps
(see also \figref{fig:sep_gen}):
\begin{enumerate}
    \item Numerical calculation of septum position using Python.
          The Python script creates a text file with the coordinates of the 
          septum strips. \opalcycl uses these coordinates to place \verb|CCOLLIMATOR|
          objects. It also generates a Visual Basic macro
          to automatically generate the 3D model in Inventor.
    \item Generation of a 3D CAD model of the septum and puller electrodes 
          in Autodesk Inventor \cite{inventor}. The macro from Step 1 is used.
    \item Import of the Inventor model into COMSOL \cite{comsol} and calculation 
          of the electrostatic fields. A mesh refinement study was performed
          to determine the correct mesh size, which was then kept throughout the 
          process.
    \item Import of the 3D field into \opalcycl as a static field map. This is achieved
          by loading it as an RF-field map with $\omega_{\RM{RF}}$ set to zero.
\end{enumerate}
In the first iteration, a baseline high-fidelity OPAL simulation without septa is used
for the placement of a septum that has twice the nominal gap width and twice the nominal voltage (thus keeping the electric field strength the same, while giving the beam space to
move radially outwards as intended). In the second iteration, the new trajectories are
used to place the septum and puller electrodes, now with nominal width and voltage, 
symmetrically around the beam in their final position. This process is then repeated for the second septum-puller pair.

\section{Beam Dynamics Simulations in the IsoDAR cyclotron}
\label{sec:main}
A preliminary study of the IsoDAR cyclotron was performed in \cite{yang:daedalus}
with encouraging results.
Since then, the harmonic was changed from 6 to 4 and the mean starting energy
was reduced from 1.5~MeV/amu to 194~keV/amu. This is a change in
starting position from the fourth turn down to the first turn.
More careful collimator placement and a full 3D treatment of the 
electrostatic septa were added to the simulations to demonstrate the 
capability of accelerating and extracting 5~mA of \htp.

\begin{figure}[t!]
    \centering
    \includegraphics[width=0.7\textwidth]{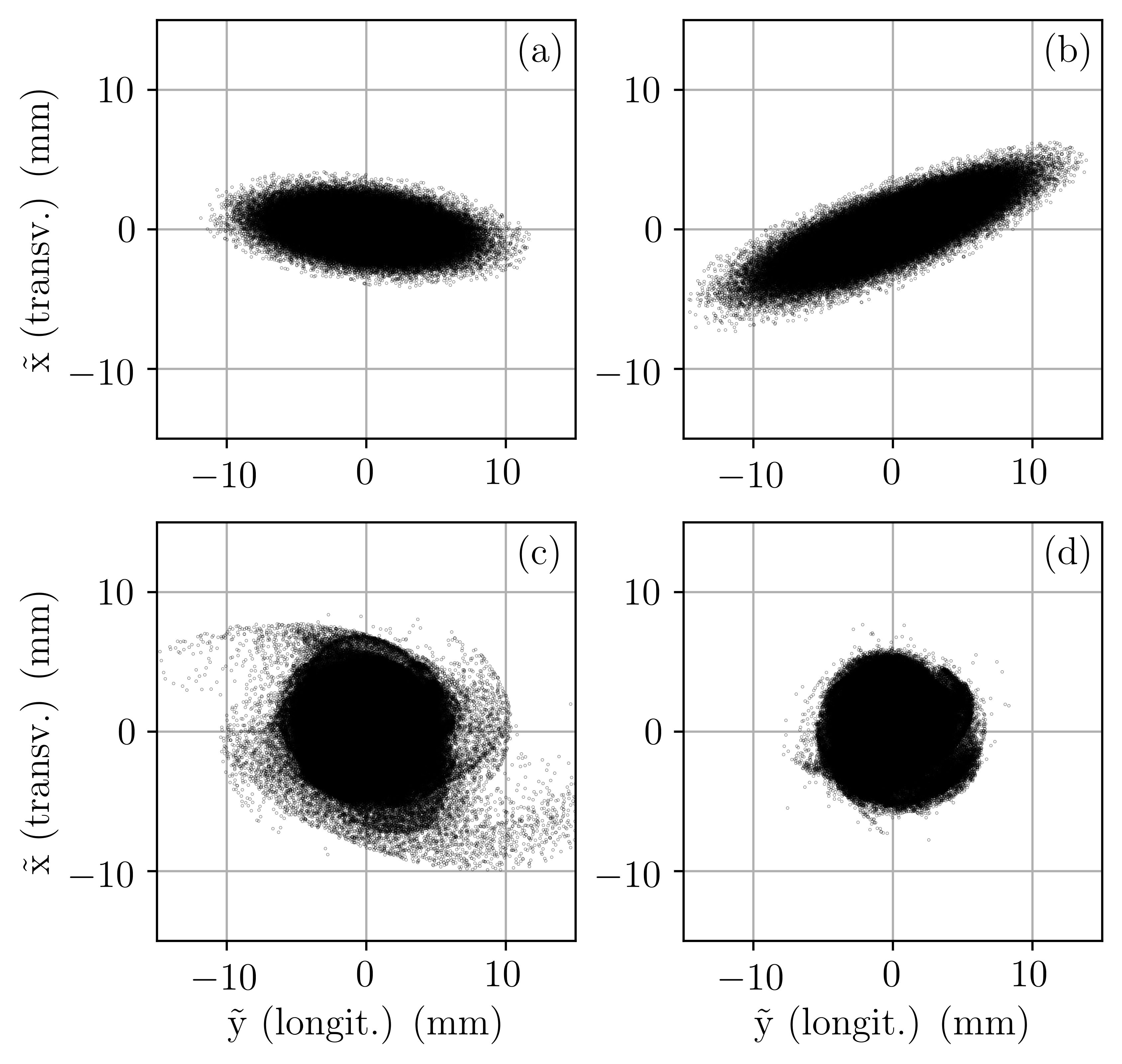}
    \caption{Beam projections onto the median plane in local frame.
             (a) Turn 0, initial beam, (b) Turn 6, no space charge.
             (c) Turn 6, $\RM{I}_\RM{beam} = 6.65$~mA, no collimators, 
             (d) Turn 6, $\RM{I}_\RM{beam} = 6.65$~mA, 12 collimators
             The development of a round $\tilde{\mathrm{x}}$-$\tilde{\mathrm{y}}$ 
             distribution can be seen
             after six turns when space charge is present (c, d). Halo that is formed in
             the process can be removed with collimators (d).}
    \label{fig:beam_distr}
\end{figure}

\subsection{First Turns and Collimation}

\begin{table}[!b]
    \setlength{\tabcolsep}{4pt}
	\vspace{-10pt}
	\footnotesize
	\caption{Initial bunch parameters in the local frame.}
	\label{tab:init_beam}
	\centering
    \vspace{5pt}
    \renewcommand{\arraystretch}{1.25}
		\begin{tabular}{ll|ll}
            \hline
            \textbf{Parameter} & \textbf{Value} & 
            \textbf{Parameter} & \textbf{Value}\\
            \hline \hline
            Distr. type & Gaussian & E$_{\mathrm{kin, mean}}$ & 194 keV/amu\\
            $\sigma_{\tilde{\mathrm{x}}}$ &  1~mm & $\tilde{\mathrm{x}}$-cutoff & $4\sigma$\\
            $\sigma_{\tilde{\mathrm{y}}}$ &  3~mm & $\tilde{\mathrm{y}}$-cutoff & $4\sigma$\\
            $\sigma_{\tilde{\mathrm{z}}}$ &  5~mm & $\tilde{\mathrm{z}}$-cutoff & $4\sigma$\\
		    $\epsilon_{\tilde{\mathrm{x}}, \RM{RMS, norm.}}$ & 0.14 mm-mrad &
		    $\epsilon_{\tilde{\mathrm{z}}, \RM{RMS, norm.}}$ & 0.59 mm-mrad \\
		    \hline
		\end{tabular}
\end{table}

\begin{figure}[t!]
    \centering
    \includegraphics[width=0.8\textwidth]{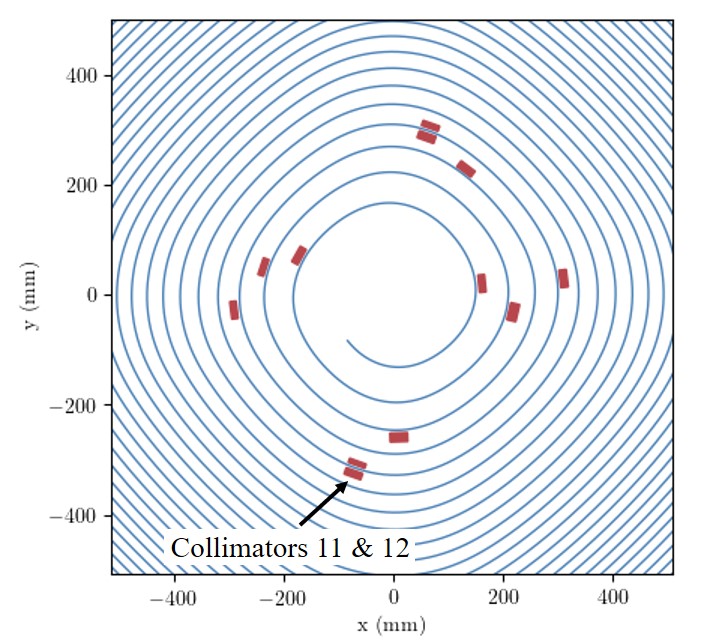}
    \caption{The centroid trajectory of the  bunch (blue) in the first turns of 
             the cyclotron after injection. 
             Twelve collimators (red) have been placed along the beam to intercept and 
             limit halo particles.}
    \label{fig:colls}
\end{figure}

\begin{figure}[b!]
    \centering
    \includegraphics[width=0.6\textwidth]{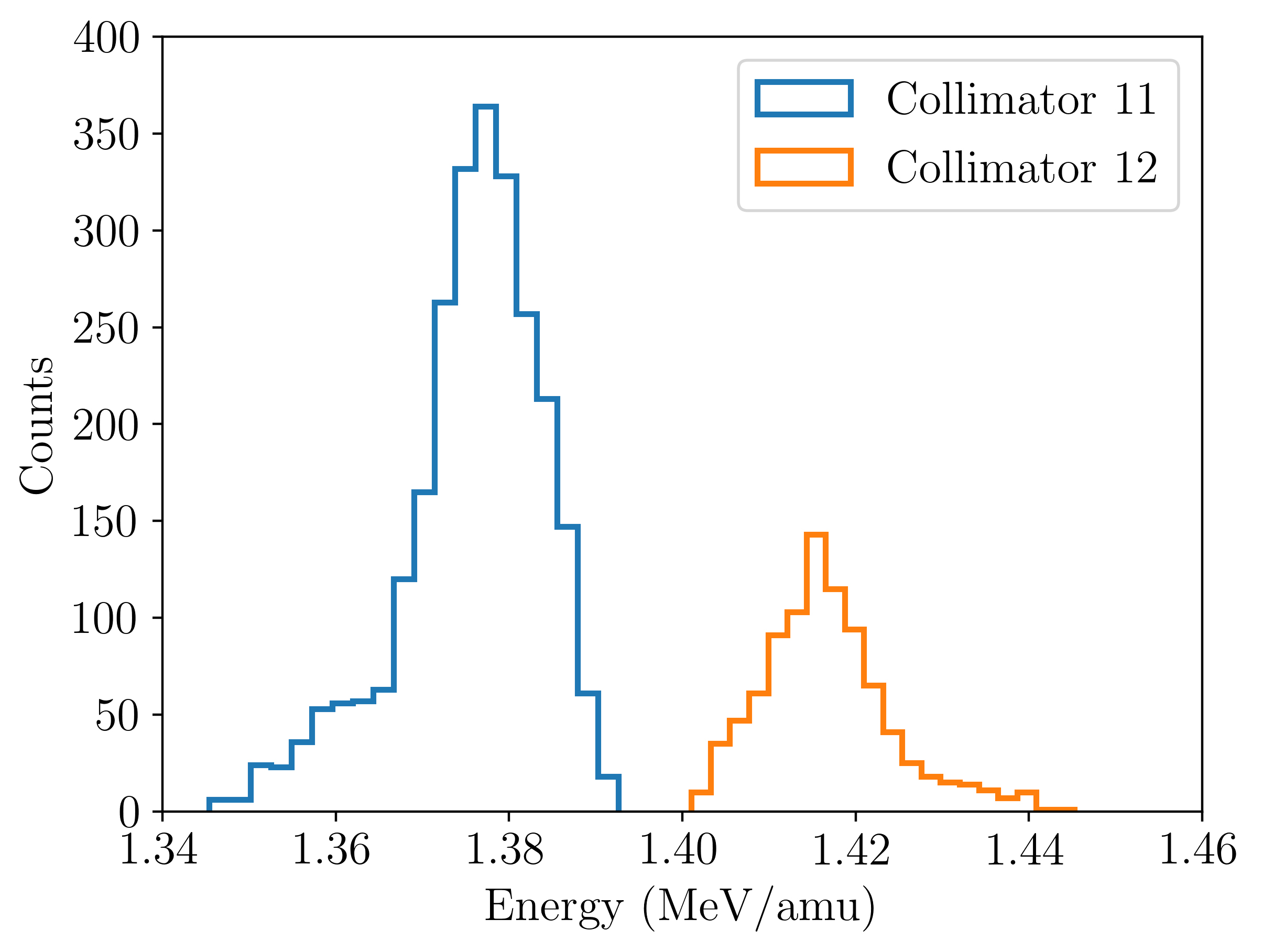}
    \caption{Energy histogram of particles lost on collimators 11 and 12. 
             Total number of accelerated particles: $10^5$.
             Collimator 11 is on the inside of the orbit, 
             Collimator 12 on the outside. Both are indicated in 
             \figref{fig:colls}.}
    \label{fig:coll_energy}
\end{figure}

The initial beam distribution is unmatched to mimic the behaviour out 
of the spiral inflector. It is
Gaussian in all three spatial directions (cutoff at $4\sigma$). 
The parameters are listed in \tabref{tab:init_beam}.
The beam is large in the vertical direction ($\tilde{\mathrm{z}}$) and has large emittance. 
A local frame projection of the beam onto the median plane ($\tilde{\mathrm{x}}$-$\tilde{\mathrm{y}}$) is 
shown in \figref{fig:beam_distr}, top, right.
After seven turns, the stationary (matched) distribution has formed as
can be seen in \figref{fig:beam_distr}, bottom-right. The halo that has
formed in the process is cut away by placing 12 collimators
(see \figref{fig:colls}). The number of particles intercepted by these
collimators and their energy is shown in the histogram in
\figref{fig:coll_energy}. It can be seen that the highest energy of terminated particles
stays below 1.5~MeV/amu. This is below the threshold for overcoming 
the Coulomb barrier and thus no activation of the collimators will
occur. The relative losses on the central region collimators are $\sim 30$\,\%.

\subsection{Acceleration}
\begin{figure}[tb]
    \centering
    \includegraphics[width=0.8\textwidth]{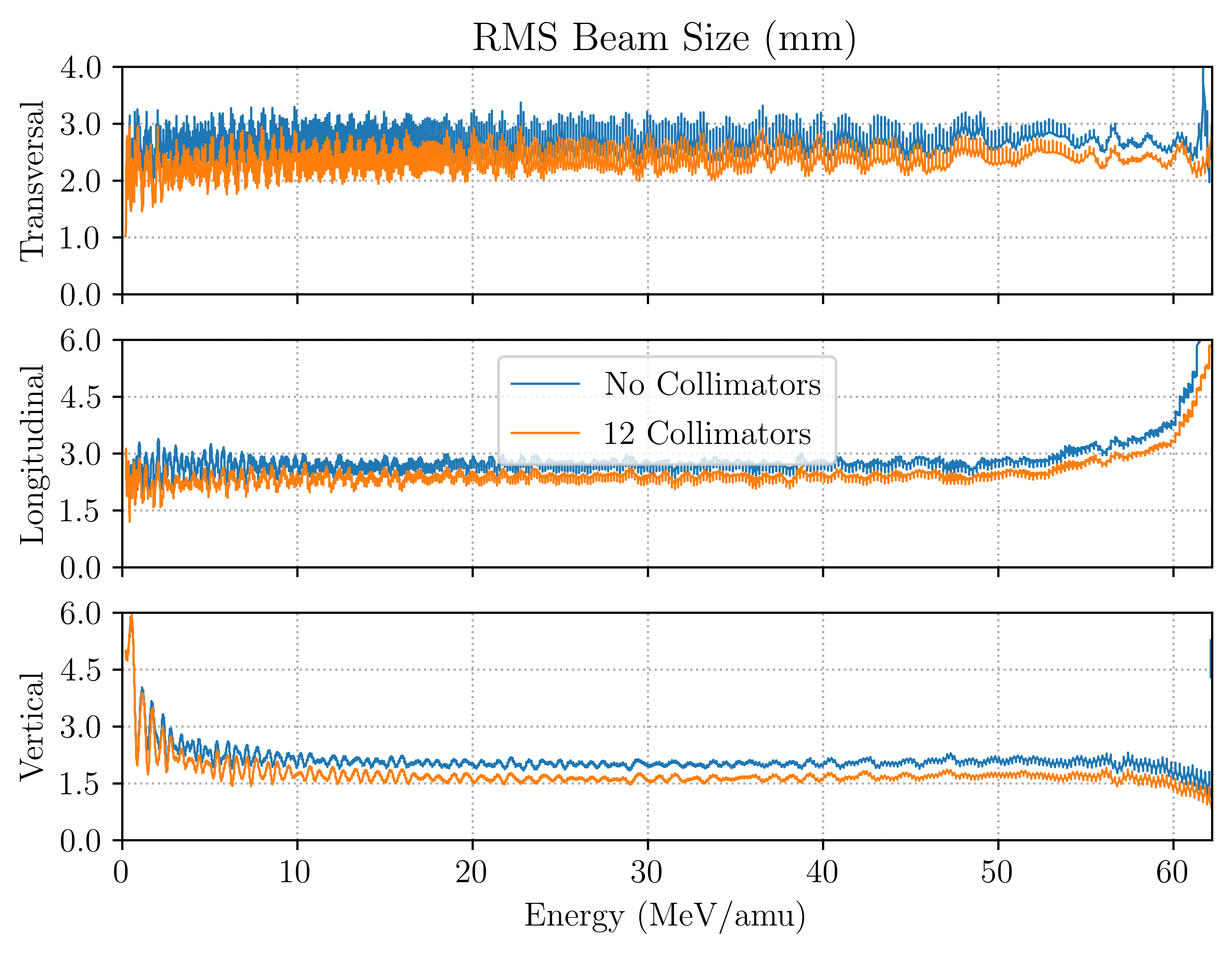}
    \caption{RMS beam size for two cases: No collimators and 12 collimators. 
             A reduction in size can be seen with collimators. 
             Also visible is the effect of the $\nu_r=1$ resonance above 60~MeV/amu.
             The longitudinal and radial beam size are approximately the same 
             above 5~MeV/amu, due to the vortex effect.}
    \label{fig:accel_size}
\end{figure}
\begin{figure}[tb]
    \centering
    \includegraphics[width=0.8\textwidth]{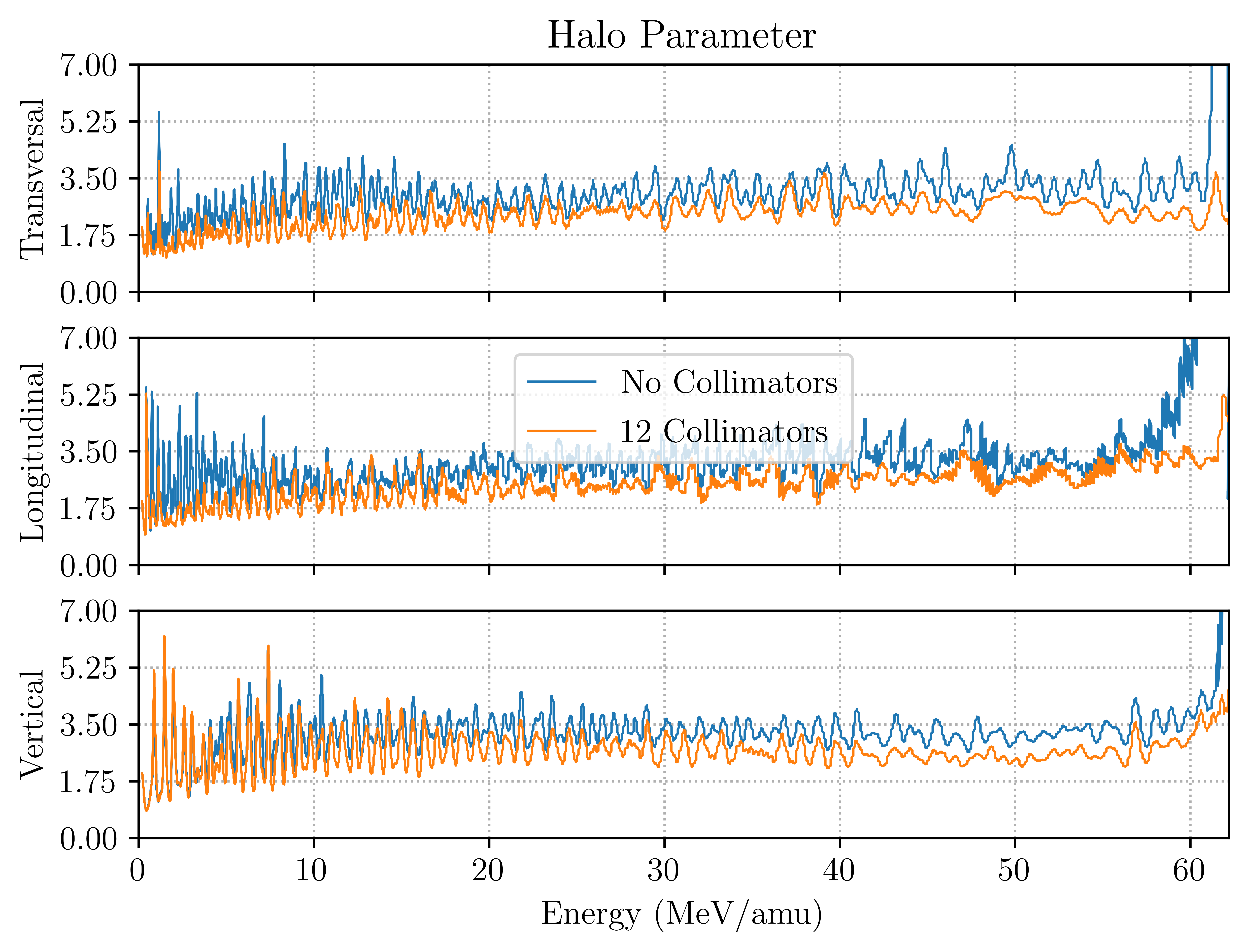}
    \caption{Halo parameter (cf. Equation \ref{eq:halo}) for two cases: 
             No collimators and 12 collimators. 
             A reduction can be seen with collimators. Also visible is the effect of the $\nu_r=1$ resonance above 60~MeV/amu. The halo
             increases significantly above 60~MeV/amu in the case without 
             collimators.}
    \label{fig:accel_halo}
\end{figure}
After the bunch has cleared the central region (10th turn), OPAL is switched into
a mode that saves full particle distributions only 4 times per turn, to not generate
an overflow of data. We then run up to 103 turns (60 MeV/amu). During the acceleration,
we use the RMS beam size and halo parameter as metrics.

The halo parameter is defined as
\BE
\label{eq:halo}
H = \frac{\langle x^4 \rangle}{\langle x^2 \rangle^2}-1
\EE

\begin{figure*}[t]
    \centering
    \includegraphics[width=1.0\textwidth]{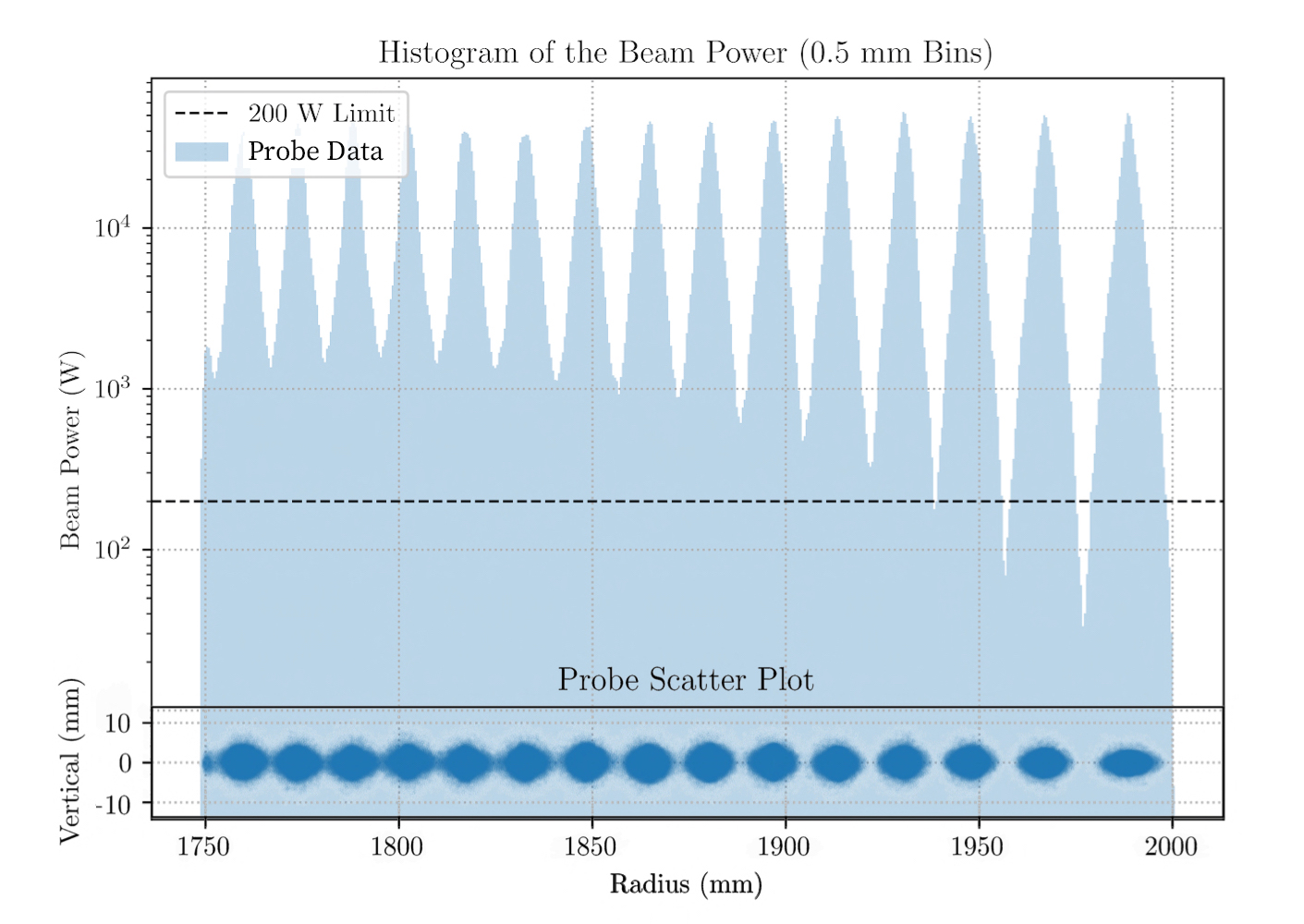}
    \caption{Probe 1 (placed at 25\degree azimuth, where 0\degree azimuth is the positive x-axis). 
            The beam power is binned in 
             0.5 mm bins (this is a conservative choice for septum width) versus 
             radius from the center of the cyclotron. R-Z scatter plot of beam spread passing through the probe has been overlaid to scale. 
             It can be seen that on a septum inserted at the appropriate radial 
             position, only about 35 W of power would be deposited.}
    \label{fig:accel_probe}
\end{figure*}

and gives an idea of the ratio of particles in a low density halo versus those
in the dense core of the bunch.
The RMS beam sizes, and halo parameter are shown in 
\figref{fig:accel_size} and \figref{fig:accel_halo}, respectively.
A clear reduction of oscillations and size can be seen with collimators. 
Also visible is the effect of the $\nu_r=1$ resonance above 60~MeV/amu.
The beam power on an \opalcycl probe (placed at 25\degree azimuth and ranging from 
$\RM{R} = 1.75$~m to $\RM{R} = 2.0$~m, where 0\degree azimuth is the positive x-axis), 
binned in 0.5 mm bins, is shown in
\figref{fig:accel_probe}. 0.5~mm is a very conservative choice for septum width,
and even so, the beam power deposited (35~W) is far below the 200~W threshold.
Beam power on a 2D probe can, of course only be an estimate of the septum losses
and in the next step, we consider a full 3D treatment of the electrostatic 
channels.

\subsection{Extraction}
\begin{figure*}[tb]
    \centering
    \includegraphics[width=1.0\textwidth]{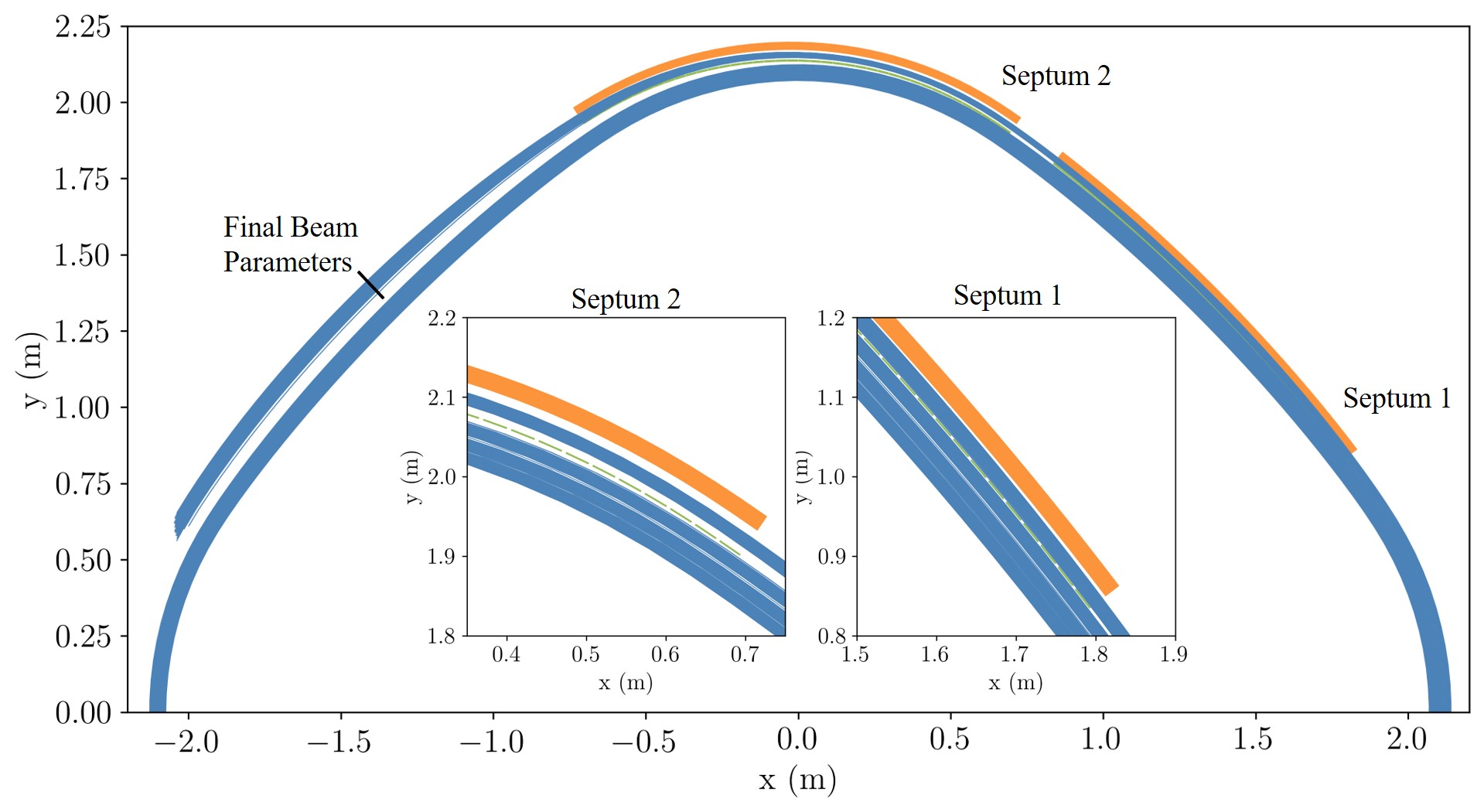}
    \caption{The final turns in the cyclotron. 1000 randomly sampled trajectories
             are displayed with the septa and puller electrodes. The position of the
             entrance to the magnetic channel at 135\degree azimuth is indicated, 
             where the turn separation is 8.5 cm center-to-center, and parameters of the final
             beam are extracted.}
    \label{fig:ext_septa}
\end{figure*}

\begin{table}[!b]
    \setlength{\tabcolsep}{4pt}
	\vspace{-10pt}
	\footnotesize
	\caption{Final bunch parameters in turn 103 at 135\degree azimuth (see 
	         \figref{fig:ext_septa}).}
	\label{tab:final_beam}
	\centering
    \vspace{5pt}
    \renewcommand{\arraystretch}{1.25}
	    \begin{tabular}{ll|ll}
            \hline
            \textbf{Parameter} & \textbf{Value} & 
            \textbf{Parameter} & \textbf{Value}\\
            \hline \hline
            E$_{\mathrm{kin, mean}}$ & 62.4 MeV/amu & 
            $\Delta$E & 0.17 MeV \\
            $\sigma_{\tilde{\mathrm{x}}, \RM{RMS}}$ &  7.5~mm &
            $\epsilon_{\tilde{\mathrm{x}}, \RM{RMS, norm.}}$ & 3.8 mm-mrad \\
            $\sigma_{\tilde{\mathrm{y}}, \RM{RMS}}$ &  11.0~mm &
            $\epsilon_{\tilde{\mathrm{y}}, \RM{RMS}}$ & 0.1 MeV-deg \\
            $\sigma_{\tilde{\mathrm{z}}, \RM{RMS}}$ &  1.9~mm & 
            $\epsilon_{\tilde{\mathrm{z}}, \RM{RMS, norm.}}$ & 0.44 mm-mrad \\
		    \hline
		\end{tabular}
\end{table}

The extraction channels are generated as described in Section \ref{sec:methods}
and the final placement can be seen in \figref{fig:ext_septa}. After careful
optimization, the combined 
beam losses on both septum electrodes are below 50~W (1e-4 relative particle 
losses). The parameters of the beam about to enter the magnetic channels are
recorded at the ``extraction point'', at an azimuthal position of 135\degree (as indicated in 
\figref{fig:ext_septa} as ``final beam parameters'') and listed in \tabref{tab:final_beam}.
All values are RMS, and normalized where applicable. 
The vertical size and emittance are small, owing to the 
vertical focusing of the isochronous cyclotron. It can be seen that the 
longitudinal and radial sizes no longer match in the way we would 
expect from vortex motion. This is due to phase slipping and
entering the $\nu_r=1$ resonance region. At the extraction point,
the turn separation is 8.5 cm center-to-center, leaving ample space for
magnetic channels.

\subsection{Beam Current Variation}
\label{sec:curr_var}
\begin{figure}[b!]
    \centering
    \includegraphics[width=0.6\textwidth]{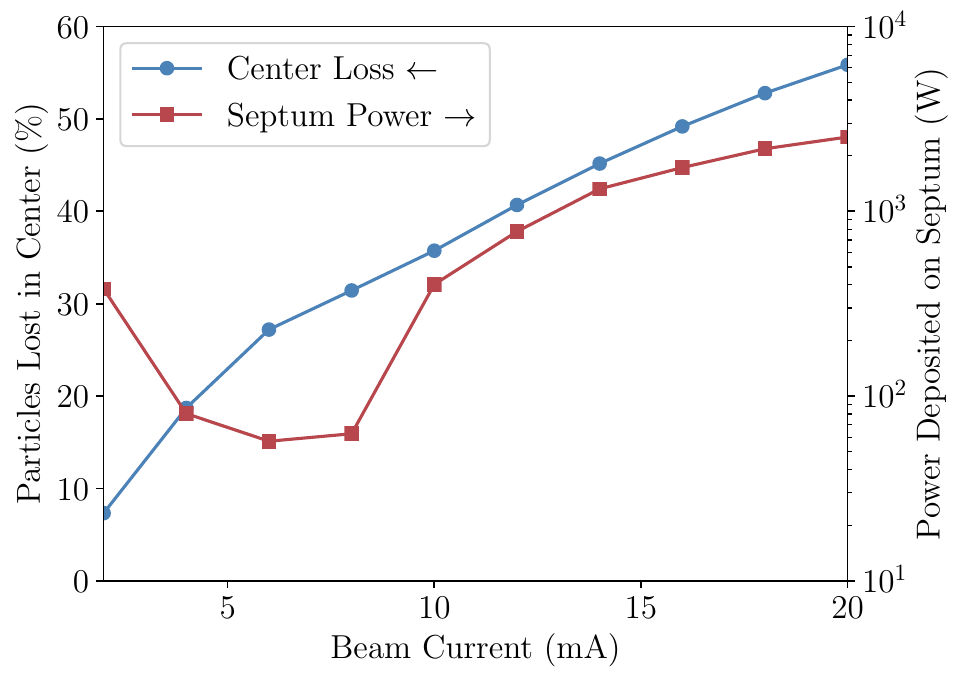}
    \caption{Losses in the central region and on the first septum versus
             initial beam currents. The placement of collimators and electrostatic
             extraction channels (septa) was optimized for 6.65~mA.}
    \label{fig:curr_var}
\end{figure}
In this part of the study, all parameters were held fixed and identical to 
the previous subsections. 
The only exception being the total beam current, which was varied 
from 2~mA to 20~mA in steps of 2~mA. No re-tuning was performed, assuming
that the collimator and electrostatic extraction channel placements are 
fixed like in a running machine. The losses in the central region and during
extraction are shown in \figref{fig:curr_var}. It can be seen that losses on the 
septa rise for low beam currents as well as high beam currents. This is discussed below.

\subsection{Discussion of Simulations}

The most important results taken from this large set of high-fidelity simulations
is that $<$50~W loss ($<$1e-4 relative loss) on the extraction septum, with good 
beam quality and excellent separation, as the beam enters the magnetic channels,
is possible for a 5~mA~\htp beam in a compact cyclotron. To achieve this, we 
incur $\sim$30\% loss on collimators below 2~MeV/amu
We note that the design of magnetic extraction channels are an engineering 
task outside of the scope of this paper. 
A preliminary study of the beam envelope
in the magnetic extraction channel was presented in \cite{calanna:isodar} 
assuming very conservative emittance values that our current results 
are far below.

Another consideration is the effect of the $\nu_r = 1$ resonance on the beam size. 
Its precession effect contributes strongly to the neccessary turn separation,
however, it can be seen in \figref{fig:accel_size} and \figref{fig:accel_halo} that, 
after 60 MeV/amu, the longitudinal beam size and halo parameter both increase strongly. 
As demonstrated, the beam quality in the last turn is sufficient for
the IsoDAR experiment, however, if more stringent restrictions have to
be placed on the beam quality, \figref{fig:accel_probe} shows that the preceding
turn also has septum losses below 200~W.

An interesting observation in Subsection \ref{sec:curr_var} is that, if the
beam current becomes too low, the relative number of particles lost on the
septum rises again and at the very low end, the beam power on the septum
becomes high enough to pose a problem. This hints at vortex motion not
being properly established if the space charge forces are too small.
Machine protection mechanisms must hence be introduced also in case of 
sudden reduction in LEBT beam current output. Similarly, pulsed beams 
must be used during commissioning rather than reduced current beams to
guarantee full space charge in each bunch. 

Everything we have observed points to clean extraction from the IsoDAR cyclotron
being possible. However, as an interesting mitigation method for high
losses on the first septum, which is only possible for \htp beams, 
the idea of a shadow foil protecting the septum was introduced
\cite{abs:isodar}. Here, a narrow carbon foil is placed in front of the 
septum electrode and \htp particles that would otherwise strike it are
now split into two protons. Due to their different magnetic rigidity, the protons 
follow a new path and can safely be extracted. In \cite{waites:isotopes}
an idea was presented to use these particles for the symbiotic production of 
radioisotopes for medical applications.

\clearpage
\section{Uncertainty Quantification}
\label{sec:uq}
In order to understand how the IsoDAR cyclotron model compares with the true physics 
behind it, uncertainty quantification techniques are used. For the high-intensity 
cyclotron design, we focus on global sensitivity analysis, which is performed to 
test how certain output Quantities of Interest (QoI), such as emittances, halo 
parameters, and RMS beam sizes, depend on the input parameters \cite{adelmann_2019}. 
This allows one to quantify error propagation in the cyclotron design, as well as 
determine its robustness.
 
\subsection{Theory}
Generally, the sensitivity of output variables to input parameters can be quantified 
through Sobol' indices \cite{sobol_2001}. These indices are obtained from an ANalysis 
Of VAriance (ANOVA) decomposition of the model's response function. It seeks to 
attribute the variability of the output to the different input parameters, while 
also taking into account correlations between them. A few mathematical bases are 
presented below:
    
Let $\vec{x}\in\mathbb{R}^d$ be the design variables, and $f(\vec{x})$ the output of the model. The ANOVA decomposition is given by:

\begin{equation*}
    f(\vec{x}) =  f_0+\sum_{i=1}^d f_i(x_i)+\sum_{1\leq i<j\leq d} f_{ij}(x_i, x_j) +
                   ... + f_{1,2,...,d}(x_1, x_2, ..., x_d),
\end{equation*}

where $\int_0^1 f_{i_1,...,i_s}(x_{i_1},...,x_{i_s}) dx_{i_k} = 0$ for $1\leq k\leq s$, and $f_0 = \int_{[0,1]^d} f(\vec{x}) d\vec{x}$ is the mean.
The total variance D can be written and decomposed as:
    
\begin{equation*}
    D = \int_{[0,1]^d}f^2(\vec{x})d\vec{x} - f_0^2 = \sum_{i=1}^d D_i + 
    \sum_{1\leq i<j\leq d} D_{ij} +
    ...+D_{1,...,d},
\end{equation*}
    
with  $D_{i_1,...,i_s} = \int_{[0,1]^s} f^2_{i_1,...,i_s}(x_{i_1},...,x_{i_s}) dx_{i_1}...dx_{i_s}$ for $1\leq i_1 < ... < i_s \leq d$. 
Then the main Sobol' indices and total Sobol' indices are given by:
    
    \begin{equation*}
        S_{i_1,...,i_s} = \frac{D_{i_1,...,i_s}}{D} \hspace{8mm} S_{i}^T = \sum_{\mathcal{I}_i} \frac{D_{i_1,...,i_s}}{D}
    \end{equation*}
    
where $\mathcal{I}_i= \{(i_1,...,i_s):\exists k, 1\leq k \leq s, i_k=i\}$. 
The main Sobol' indices quantify the effect of a parameter on the output variable 
without taking into account correlations with other parameters. The total Sobol' 
indices do take these into account, and are the most important ones for a global 
sensitivity analysis \cite{frey_2021}.
    
The Sobol' indices can be computed via Monte-Carlo simulations. However, 
due to the computational costs of these types of simulations, other less 
expensive techniques must be explored. One such technique is to use 
Polynomial Chaos Expansion (PCE). A short theoretical overview is given 
below \cite{adelmann_2019}.
    
Let $(\Omega, \mathcal{F}, \mathcal{P})$ be a complete probability space. 
The design variables can be written as random variables $\vec{x}\in \mathbb{R}^d$, 
where $d$ is the number of design variables. The joint probability density function 
is then given by $\rho(\vec{x})=\prod_{k=1}^{d} \rho(x_k)$, where $\rho(x_k)$ is the individual probability density of the $k$-th design variable. We define also the 
set of multi-indices $\mathcal{I}_{d,p} = \{\vec{i}=(i_1,...,i_d)\in \mathcal{N}_0^d : ||\vec{i}||_1 \leq p \}$, where $p$ is the order at which we will truncate the 
polynomial. Then $\forall u(\vec{x})\in L_2(\Omega, \mathcal{F}, \mathcal{P})$, 
which corresponds to a QoI, can be decomposed as:
    
\begin{equation}
    u(\vec{x}) = \sum_{\vec{i}\in I_{d,\infty}} \alpha_{\vec{i}} \psi_{\vec{i}}(\vec{x}), 
\end{equation}
    
where $\psi_{\vec{i}}(\vec{x}) = \prod_{k=1}^{d} \psi_{i_k}(x_k)$ are the
multivariate polynomial chaos basis functions. They are obtained as the product of
$\psi_{i_k}$, the univariate polynomials of degree $i_k\in\mathcal{N}_0$, which 
satisfy the orthogonality relation $<\psi_{i_k}\psi_{j_k}> = \int_{\Omega}
\psi_{i_k}\psi_{j_k} \rho(x_k) dx_k = \delta_{i_k j_k}\mathcal{E}[\psi_{i_k}^2]$. 
The explicit form of this basis depends on the probability density function. 
The PCE approximates the QoI $u(\vec{x})$ by a truncated series 
$\hat{u}(\vec{x}) = \sum_{\vec{i}\in I_{d,p}} \alpha_{\vec{i}} \psi_{\vec{i}}(\vec{x})$.
    
Building a PCE model requires a set of high-fidelity simulation samples to train it. 
However, once this is obtained, the Sobol' indices are analytically calculated from 
the PCE coefficients by gathering the polynomial decomposition into terms with the 
same parameter dependence to obtain the ANOVA decomposition. The Sobol' indices 
follow without needing to perform more simulations, reducing computational cost
\cite{sudret_2008}.
    
Furthermore, the PCE model is now a surrogate model, i.e. a black-box that mimics 
the behaviour of the high-fidelity simulations when given a set of input parameters.
Furthermore, this surrogate model has the corollary of allowing for fast multi-objective
optimisation of the design \cite{adelmann_2019}.

\subsection{The IsoDAR Case}
\begin{figure}[ht]
    \centering
        \includegraphics[width=0.8\textwidth]{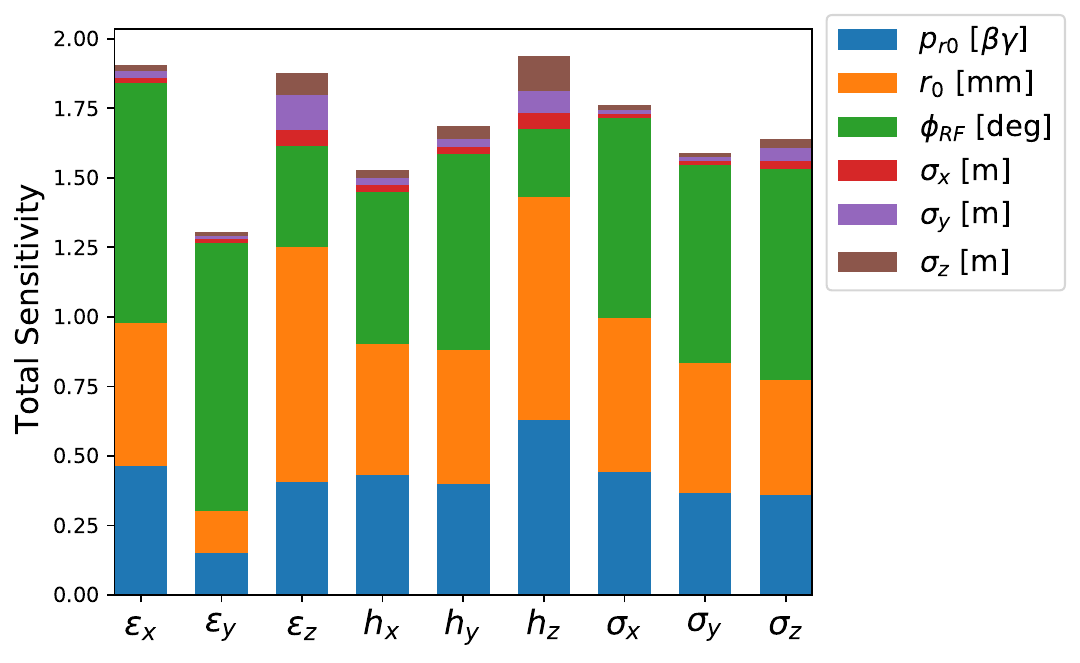}
        \caption{Global sensitivity analysis using an order 6 PCE for the 
                 IsoDAR cyclotron.\label{sens6}}
    \end{figure}

The PCE models will be constructed using the Uncertainty Quantification Toolkit (UQTk)
\cite{DebusschereUQTk_2017}. For this specific case, the polynomials will be Legendre
polynomials as we assume uniform distribution of the QoIs. The design
parameters for the IsoDAR cyclotron, as well as the 
lower and upper bounds of variation around the design value, are given in Table \ref{dvars}.
\begin{table}
\centering
    \begin{tabular}{|c|cccccc|}
        \hline
        {} & $p_{r0}$ [$\beta\gamma$] & $r_0$ [mm] & $\phi_{rf}$ [deg] & 
        $\sigma_x$ [m] & $\sigma_y$ [m] & $\sigma_z$ [m]\\
        \hline
        Lower &  0.00225 &  115.9 &  283.0 &  0.00095 &  0.00285 &  0.00475 \\
        Upper &  0.00235 &  119.9 &  287.0 &  0.00105 &  0.00315 &  0.00525 \\
        \hline
    \end{tabular}
    \caption{Design parameters that are to be varied around the 
             design values in order to perform an uncertainty
             quantification. $p_{r0}$ is the radial momentum at 
             injection \cite{OPAL-Manual}, $r_0$ is the radial position 
             at injection, $\phi_{RF}$ is the RF angle, and 
             $\sigma_{x,y,z}$ are the RMS beam sizes in each direction.} 
    \label{dvars}
\end{table}
The QoIs are measured at the 95th turn of the cyclotron. Measurement at the 103rd turn is avoided since at that point particles are artificially removed in the simulation. The QoIs are listed below.
\begin{itemize}
    \setlength\itemsep{0.05em}
    \item Projected emittances $\epsilon_{x,y,z}$ [mm-mrad]
    \item Halo parameters $h_{x,y,z}$ [-]
    \item RMS beam sizes $\sigma_{x,y,z}$ [m]
\end{itemize}
The PCE models for each quantity of interest are trained using 80\% of a 7500-point sample,
and validated on the other 20\,\%. Some oversampling was done in order to improve the fit 
in sparsely populated regions of the output. Figure \ref{sens6} shows the global 
sensitivity analysis, obtained using an order 6 model. As can be seen, the RMS beam 
sizes at injection have little to no effect on the output quantities if they vary 5\,\% 
around their design value. The RF angle $\phi_{RF}$ is the most significant design 
variable, followed by the initial radial momentum $p_{r0}$. This is consistent with 
the physics of the accelerator. The beam velocity needs to be matched to the RF phase 
of the cavity to ensure the beam is accelerated and focused, so it is to be expected 
that $p_{r0}$ and $\phi_{RF}$ have the most impact on final beam properties. The 
injection radius $r_0$ is also important, since this is a parameter which should be 
precisely set so that the beam does not arrive at an undue time at the accelerating 
cavity. The robustness of the model is ensured by realizing that most significant 
design variables are fully controllable. It is also corroborated by the physical 
consistency of the model.
    
\subsection{Surrogate Model}
\begin{figure*}[!t]
    \includegraphics[width=0.32\textwidth]{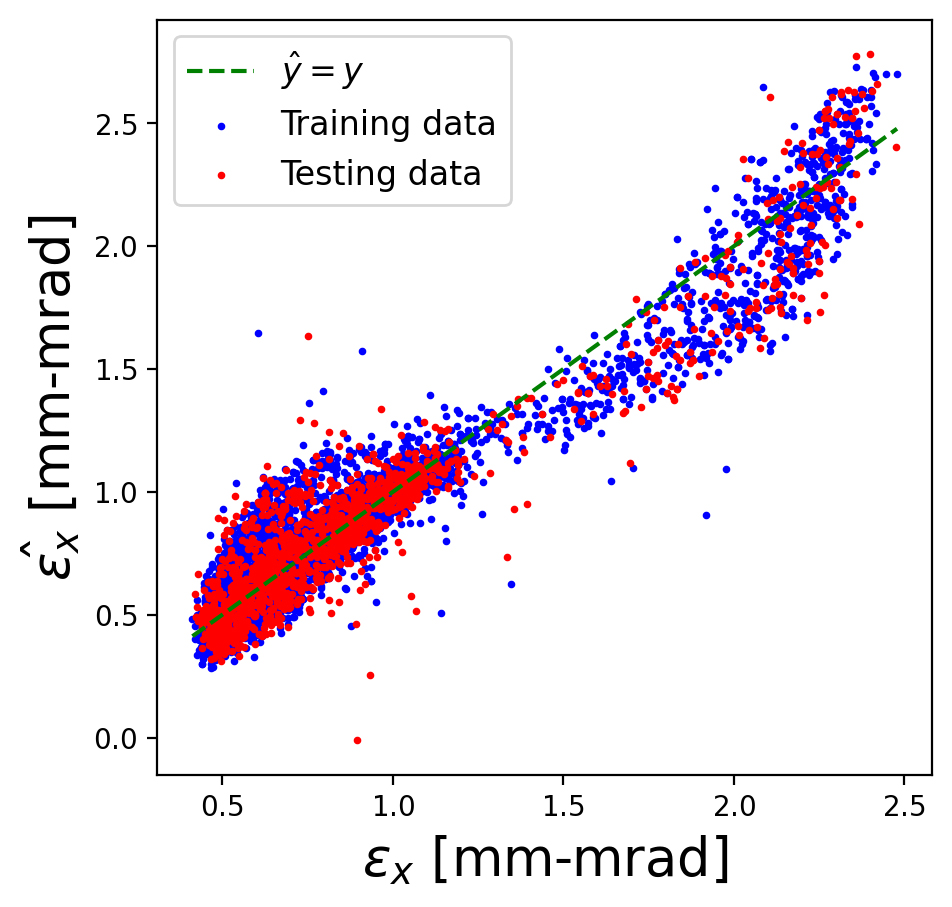}
    \includegraphics[width=0.32\textwidth]{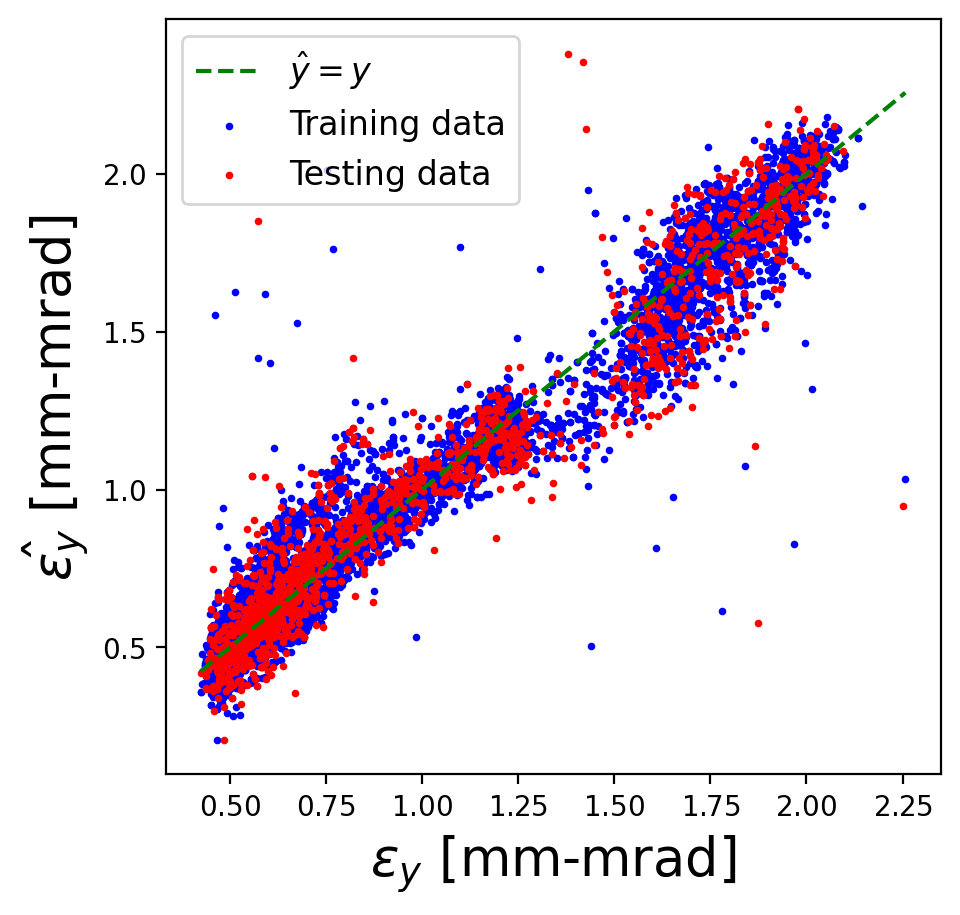}
    \includegraphics[width=0.32\textwidth]{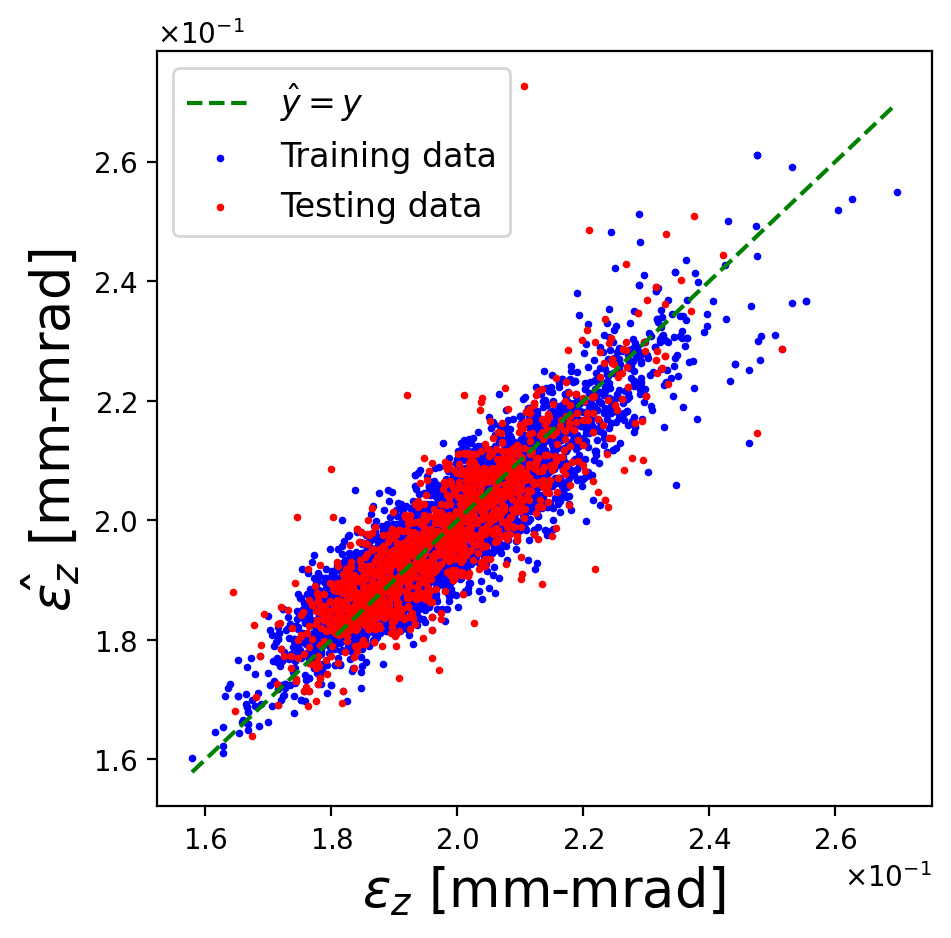}
    \caption{Surrogate model predicted value (indicated with a hat) versus OPAL 
             simulation value of the projected emittances in all three planes, 
             for training and testing points. The PCE model would perfectly 
             replicate the high-fidelity simulations if all the points were 
             lying on the 45\degree dashed line.\label{emit}}
        
\end{figure*}

\begin{figure*}[!t]
    \includegraphics[width=0.32\textwidth]{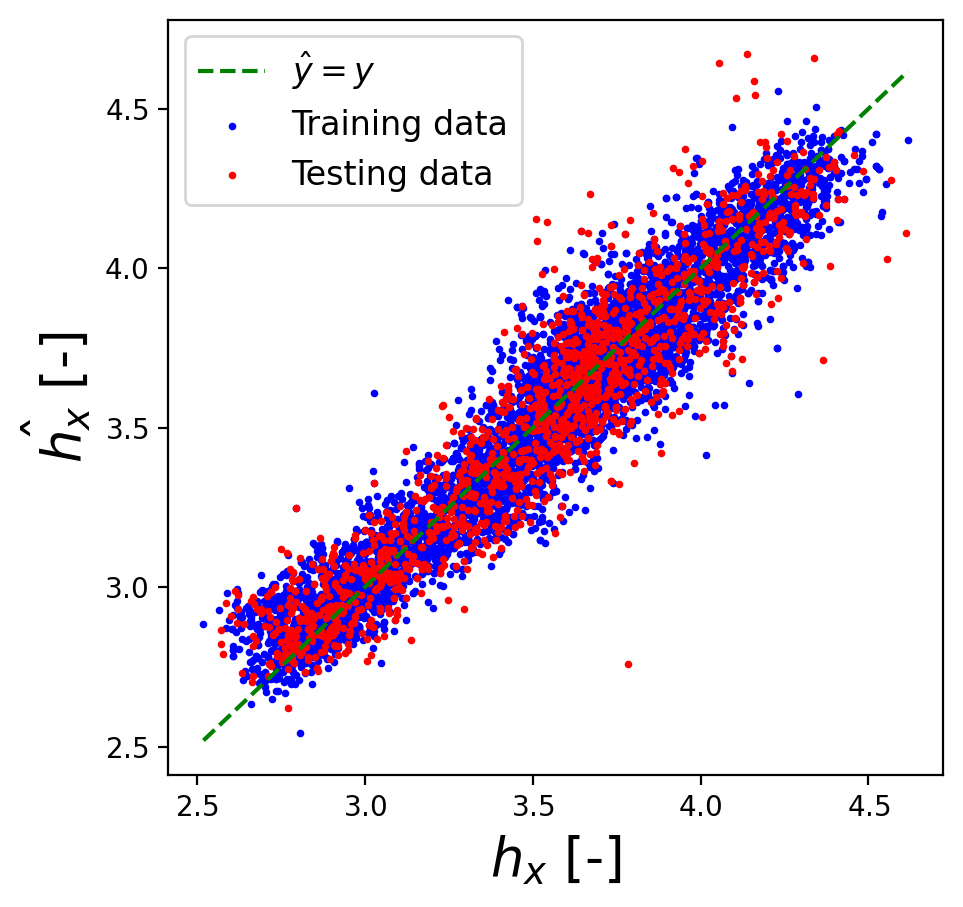}
    \includegraphics[width=0.32\textwidth]{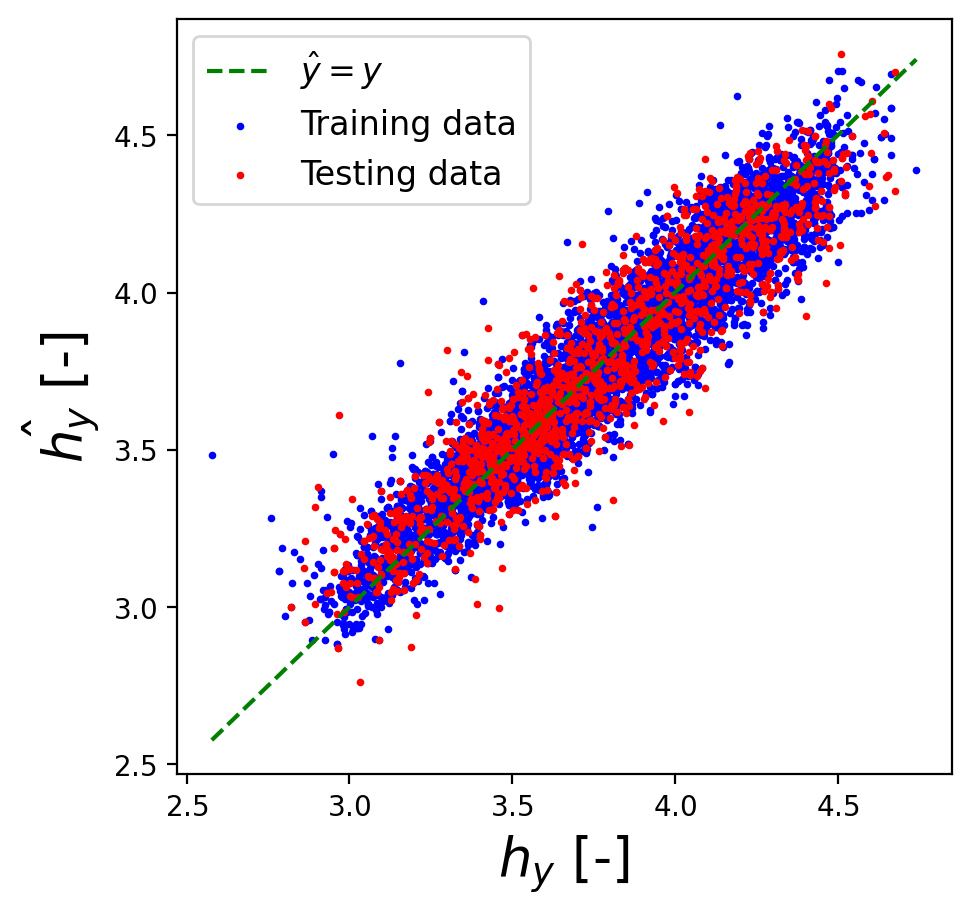}
    \includegraphics[width=0.32\textwidth]{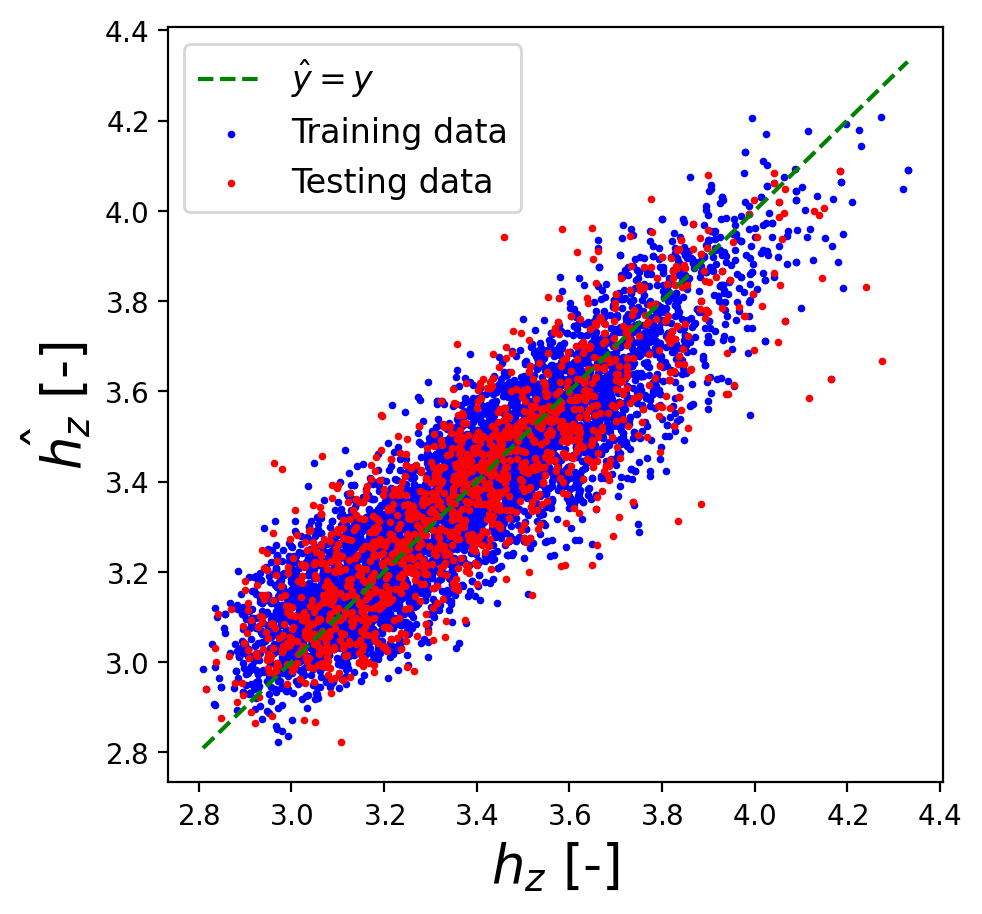}
    \caption{Surrogate model predicted value (indicated with a hat) versus OPAL simulation 
             value of the halo parameters in all three planes, for training and 
             validation points. The PCE model would perfectly replicate the 
             high-fidelity simulations if all the points were lying on the 
             45\degree dashed line.\label{halo}}
\end{figure*}

\begin{figure*}[!t]
    \includegraphics[width=0.32\textwidth]{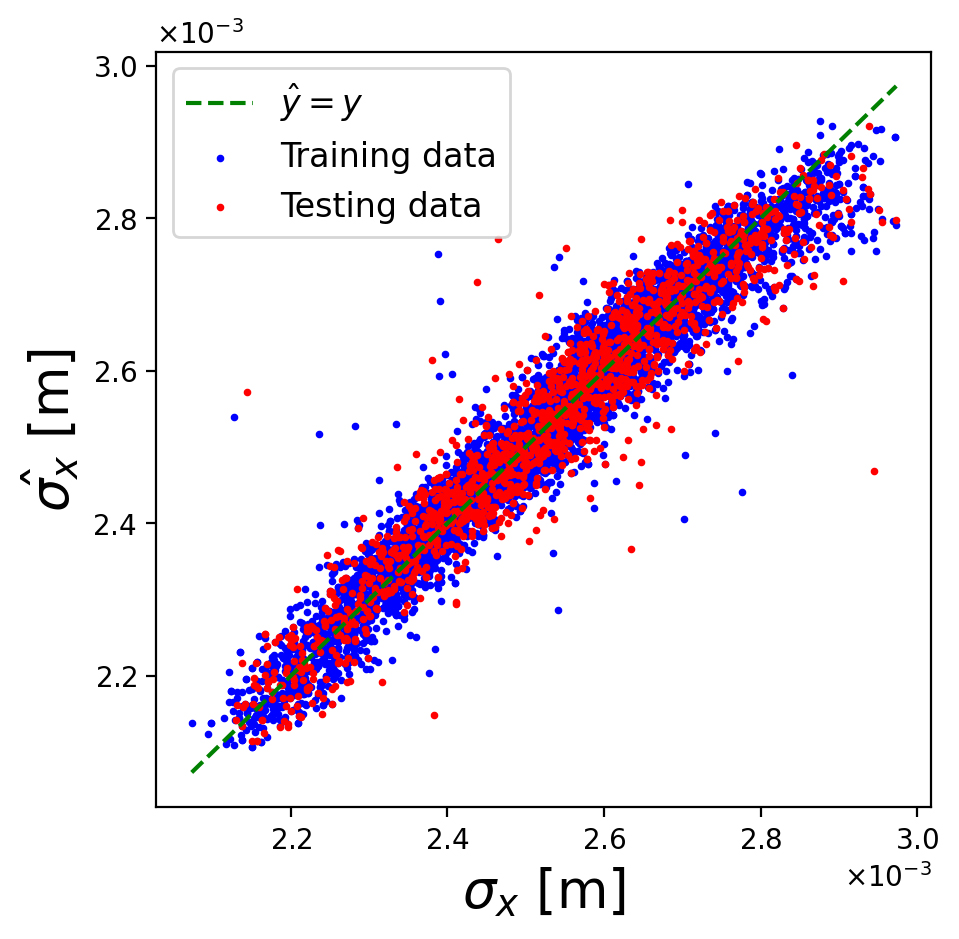}
    \includegraphics[width=0.32\textwidth]{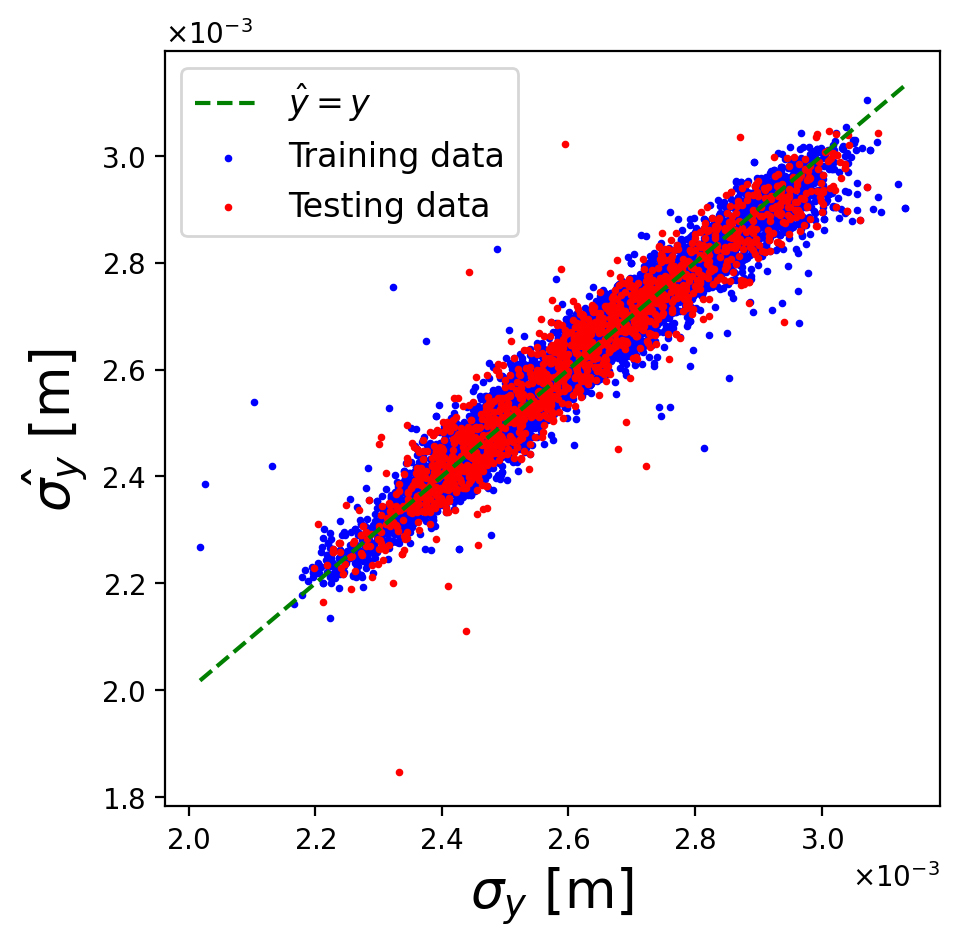}
    \includegraphics[width=0.32\textwidth]{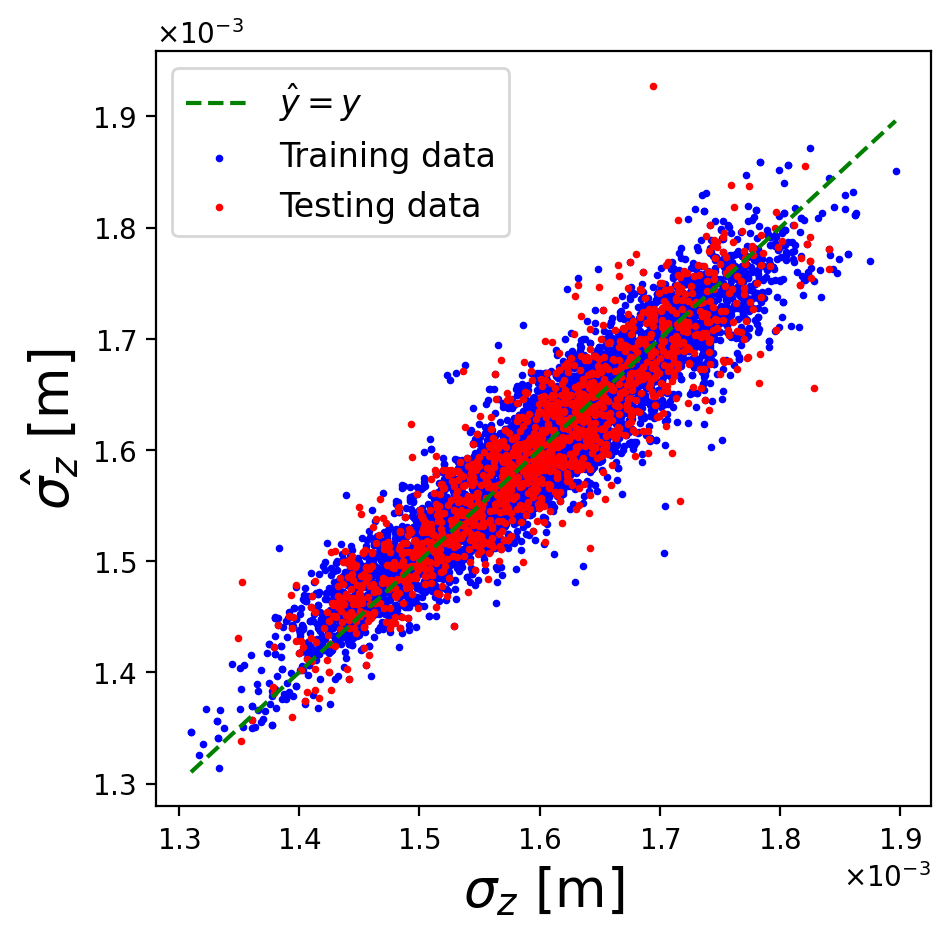}
    \caption{Surrogate model predicted value (indicated with a hat) versus OPAL 
             simulation value of the RMS beam sizes in all three planes, for 
             training and validation points. The PCE model would perfectly 
             replicate the high-fidelity simulations if all the points were 
             lying on the 45\degree dashed line.\label{rms}}
\end{figure*}

The reliability of the surrogate model is seen by comparing the values of the QoIs obtained through the high-fidelity OPAL simulations versus those predicted by 
the PCE model. This is shown in Figures \ref{emit}, \ref{halo}, and \ref{rms}, for the 
order 6 model.
    
The models for the emittances exhibit a specific pattern which is not yet understood yet, but they stay close to the $\hat{y}=y$ line nonetheless. Some anomalies show departures from the main trend, but generally the predictions correspond well to the values from the simulations. 
The halo parameters and RMS beam sizes have better predictions. The Mean Absolute 
Errors (MAE) on the training and the validation set can be found 
in Table \ref{errors}.
The MAE test and train errors stay below 5\% for all QoIs except 
for the projected emittances on the $x$-plane and the $y$-plane, $\epsilon_x$ 
and $\epsilon_y$. This can be attributed to the emittance being a quantity that is 
generally hard to compute. In our experiments we found that the order 6 PCE model proved to minimize the MAE for the testing set. Increasing order more than 6 caused the model to over-fit on the training set. Overall, the errors are reasonable, and \figref{emit}, \figref{halo}, and \figref{rms} show a good fit between high-fidelity and surrogate model values.

\begin{table}[b!]
    \centering
    \caption{Mean Absolute Error (MAE) of the surrogate models in percentage for the training 
             and the testing sets.\label{errors}}
    \begin{tabular}{lrr}
        \hline
        {} &  MAE train [\%] &   MAE test  [\%] \\
        \hline
        $\epsilon_x$ [mm-mrad] &   9.397187 &  11.643873 \\
        $\epsilon_y$ [mm-mrad] &   7.621135 &   9.097649 \\
        $\epsilon_z$ [mm-mrad] &   2.065314 &   2.378235 \\
        $h_x$ [-]                    &   2.776718 &   3.372337 \\
        $h_y$  [-]                   &   2.477438 &   2.968650 \\
        $h_z$ [-]                    &   2.635573 &   3.027616 \\
        $\sigma_x$ [m]               &   1.221384 &   1.521169 \\
        $\sigma_y$ [m]               &   1.250924 &   1.521905 \\
        $\sigma_z$ [m]               &   1.571111 &   1.760405 \\
        \hline
    \end{tabular}
\end{table}

\subsection{Discussion of Uncertainty Quantification} 
The IsoDAR cyclotron uncertainty quantification shows that the computational model 
and the physics model are consistent with each other, and gives credibility to the 
design. Furthermore, the advantage of surrogate modeling is that we obtain a 
black-box that reasonably predicts the output of a costly high-fidelity simulation 
given certain design variables at a fraction of the computational cost. These 
surrogate models are orders of magnitude faster \cite{adelmann_2019} than the OPAL
simulation. This fact can be exploited in order to perform fast multi-objective 
optimisation, for example using a genetic algorithm \cite{adelmann-2020-1}. This 
could be used to finding other optimal working points of the IsoDAR cyclotron in 
future studies. A first trial at finding another optimal working point within the
bounds presented in Table \ref{dvars} makes the optimization algorithm fall back 
to the original design values of the cyclotron, ensuring that it is indeed an 
optimum, and again verifying the robustness of the design.

\clearpage
\section{Conclusion and Outlook\label{sec:conclusion}}
In this paper, we presented a mature design, and
simulations thereof, for the IsoDAR 60~MeV/amu 
compact isochronous cyclotron, which accelerates 5~mA of \htp. The molecular
hydrogen ions can then be charge-stripped with a carbon foil, 
yielding 10~mA of protons. 
The primary application of this machine is a definitive search
for sterile neutrinos, however, the applications in other areas of science and industry 
are numerous: Material research, isotope production, energy research, and 
CP-violation searches in the neutrino sector (the latter two
when the IsoDAR cyclotron is used as an injector to a larger cyclotron).

In order to verify our design, an exhaustive simulation study, using the well-established
particle-in-cell code OPAL with 1e5 to 1e6 particles per bunch, was performed. 
Space-charge was taken into account, as well as all external fields and 
termination of particles inside the cyclotron. The extraction channels were 
modeled in CAD software and 3D fields were imported into OPAL.
Through the combined forces of the cyclotron magnet, the accelerating RF 
cavities, and the particles' self-fields, a \emph{vortex-effect} takes place, which 
we exploited to stabilize the bunch size and phase space in the longitudinal-radial
plane. This led to clean extraction, when using a set of two electrostatic channels,
where power deposition at the highest particle energies was kept below 50~W 
(a quarter of the 200~W safety limit established at PSI). This is 
sufficient to guarantee low activation of the cyclotron and hence allows for
frequent hands-on maintenance. To our knowledge, this is the first particle accelerator actively designed to exploit the vortex effect to transport and 
accelerate high intensity beams.

In the presented study, we started the design and simulation process with 
the bunch already injected into the cyclotron and coasting at 193~keV/amu.
Our ongoing work on radiofrequency-direct injection (RFQ-DIP), and a preliminary
design of the cyclotron spiral inflector and central region
(which we briefly described in Section \ref{sec:design}) give us confidence
that the particle distributions can be matched at that point and that
the total losses from ion source extraction to cyclotron extraction 
are below 50\,\%, requiring only 10~mA of DC \htp beam from the ion source.
A full start-to-end simulation of ion source, low energy beam transport,
RFQ-buncher and central region, including space-charge in all parts of the line, 
is currently ongoing and a publication is forthcoming.

We have also presented a full uncertainty quantification (UQ) using machine learning
(surrogate modeling with polynomial chaos expansion) to determine how sensitive 
our optimized design is with respect to variations of the beam input parameters. From the results we can conclude that small variations of input beam parameters within the expected limits 
can be tolerated according to the UQ and thus our design is robust. Furthermore the computational model is shown to be consistent with the physics.

In addition, we propose a novel method to protect the extraction channel,
in which \htp particles that would hit the septum are broken up into protons 
by means of a narrow stripper foil placed upstream of the septum.  
These protons will be bent inside of the septum, and follow a trajectory 
that takes them safely outside the cyclotron into either a beam dump, or a 
medical isotope target. Having this option 
is a direct consequence of the novel concept of using \htp for acceleration.

Other future work also includes multi-bunch simulations, wherein OPAL
injects five bunches in sequence (one per full turn for five turns) to account for the
space charge effect of neighboring bunches. This was done for an earlier iteration
of our cyclotron design and the results were not dramatically changed. If anything,
they were slightly improved when neighboring bunches ``pushed'' against each other 
through space charge.

\section*{Acknowledgements}
The authors are very thankful to Jose Alonso (LBNL), William Barletta (MIT), 
and Luciano Calabretta (INFN) for their past work on IsoDAR, their support, and for
fruitful discussions. 
We also thank Anastasiya Bershanska (MIT) for supporting software and 
Larry Bartoszek for help with CAD modeling. 
Furthermore we would like to thank the OPAL team for their support.
This work was supported by NSF grants PHY-1505858 and 
PHY-1626069. Winklehner was supported by funding from the Bose Foundation. Muralikrishnan has received funding from the European Union's Horizon 2020 research and innovation program under the Marie Sk{\l}odowska-Curie grant agreement No. 701647 and from the United States National Science Foundation under Grant No. PHY-1820852. 

\section*{References}
\bibliographystyle{iopart-num}
\bibliography{bibliography}

\end{document}